\documentclass[11pt,onecolumn]{IEEEtran}
\IEEEoverridecommandlockouts

\usepackage{amssymb}
\usepackage{amsfonts}
\usepackage{graphicx}
\usepackage{textcomp}
\usepackage{xcolor}

\usepackage[cmex10]{amsmath}
\usepackage{enumerate}
\usepackage{amsthm}
\usepackage{latexsym}
\usepackage{xspace}
\usepackage{amscd}
\usepackage{epic}

\def\BibTeX{{\rm B\kern-.05em{\sc i\kern-.025em b}\kern-.08em
    T\kern-.1667em\lower.7ex\hbox{E}\kern-.125emX}}

\newcommand{\N}{{\mathbb N}}
\newcommand{\Z}{{\mathbb Z}}
\newcommand{\F}{\mathbb F}
\newcommand{\X}{\mathbf {X}}
\newcommand{\fF}{\mathfrak F}
\renewcommand{\bf}[1]{\mathbf{#1}}
\newcommand{\Le}{\mathbb L}
\renewcommand{\L}{\mathbb L}
\newcommand{\A}{{\mathcal A}}

\newcommand{\D}{{\mathcal D}}
\newcommand{\R}{{\mathcal R}}

\newcommand{\bL}{\mathbf {\Lambda}}
\renewcommand{\S}{{\mathcal S}}

\newcommand{\tq}{\, \mid \,}
 
\newcommand{\supp}{{\rm supp}}

\newtheorem{theorem}{Theorem}
\newtheorem{lemma}[theorem]{Lemma}
\newtheorem{proposition}[theorem]{Proposition}
\newtheorem{definition}[theorem]{Definition}
\newtheorem{definitions}[theorem]{Definitions}
\newtheorem{corollary}[theorem]{Corollary}
\newtheorem{remark}[theorem]{Remark}

\newtheorem{example}[theorem]{Example}

\newtheorem{procedure}{Procedure}
\newtheorem*{algorithm*}{Algorithm}
    
\begin{document}

\title{A new approach to the Berlekamp-Massey-Sakata Algorithm.  Improving Locator Decoding
\thanks{\textsuperscript{*}This work was partially supported by MINECO, project MTM2016-77445-P, and Fundaci\'{o}n S\'{e}neca of Murcia, project 19880/GERM/15.}
}

\author{
\IEEEauthorblockN{Jos\'e Joaqu\'{i}n Bernal \textit{and} Juan Jacobo Sim\'on}

\IEEEauthorblockA{\textit{Departamento de Matem\'aticas} \\
\textit{Universidad de Murcia}\\
30100 Espinardo, Murcia\\
josejoaquin.bernal@um.es \textit{and} jsimon@um.es}
}

\maketitle 
\begin{abstract}
We study the problem of the computation of Groebner basis for the ideal of linear recurring relations of a doubly periodic array. We find a set of indexes such that, along with some conditions, guarantees that the set of polynomials obtained at the last iteration in the Berlekamp-Massey-Sakata algorithm is exactly a Groebner basis for the mentioned ideal. Then, we apply these results to improve locator decoding in abelian codes.
\end{abstract}


%

\section{Introduction}

%
%
%
%
\IEEEPARstart{T}{he} Berlekamp-Massey-Sakata algorihtm (BMSa for short) is a two dimensional generalization of the Berlekamp- Massey algorithm that applies (among others) to locator decoding of cyclic codes and some codes on curves. There exist many decoding methods based on the development of locator decoding in several different directions, e. g. in the context of bivariate abelian codes (see \cite{Blah,Sakata}) or in the context of algebraic-geometry codes  (see \cite{Cox et al Using,saints heegard,Sakata 5,Sakata 3}). Hence, it is of interest to improve the general locator decoding procedure \textit{per se}. On the other hand, there are another applications to construction techniques of some type of Groebner basis (see, for example \cite{bertho,fuagegre}).

The origins of this topic may be found in the computation of Groebner basis for the ideal, denoted by $\bL(U)$, of the so called linear recurring relations of a doubly periodic array, $U$ (see notation below); among others, see \cite{Imai}, \cite{Sakata 4} and \cite{rubio 2}. The BMSa appears for the first time in the paper by S. Sakata \cite{Sakata 2}. It is an iterative procedure, with respect to a well-ordering on $\N\times\N$, to construct Groebner basis. Its nature drives us to a fundamental problem: find a minimal (or maximal) number of steps or, from another point of view,  a suitable subset of $U$ with minimal cardinality to get the mentioned Groebner basis.

In \cite{Cox et al Using,Hackl,rubio} and \cite{Sakata 2} one may find some termination criteria based on the computation of bounds to that number of steps that depend on the analysis of the shape of all possible sets $\Delta(U)$ (see Definition~\ref{base de u}). In \cite{Cox et al Using,Sakata 5} and \cite{Sakata 6} some inference procedures (specially the Feng-Rao Majority Voting \cite[Theorem 10.3.7]{Cox et al Using} in the case of one point AG-codes) are applied to compute those unknown syndromes needed to complete the implementation of the BMSa. Roughly speaking, a common idea is to consider enough steps to ensure that the size of the set determined by the footprint of $\bL(U)$ (see Paragraph \ref{parrafo de conjuntos delta y puntos de def}) cannot be increased. On the other hand, in Blahut's paper \cite[p. 1614]{Blah} (see also \cite{saints heegard}), one may glimpse a new strategy to address the problem. The idea is to find a set of indexes, say $\A$, satisfying the following property: for any $l\not\in \A$, the $l$-iteration does not contribute to the constructive procedure and so, one only has to consider those iterations of elements in $\A$. Although Blahut's paper does not provide us with a set of indexes having the property described above, it stays close to achieve it. 

The goal of this paper is to find a set of indexes, that we call $\S(t)$ (see Definition~\ref{def conjunto adecuado}), such that, along with some conditions, it guarantees that the set of polynomials obtained at most at the last iteration in the BMSa over such a set is exactly a Groebner basis for $\bL(U)$ (theorems~\ref{condicion suficiente} and \ref{condicion suficiente no locator}). Moreover, in the context of locator decoding, we prove that, given a list of values in a doubly periodic array $U$ afforded by the evaluation of a (error) polynomial $e$, from a sublist of those values indexed by a set of the form $\S(t)$, one may recover both of them, the array and the polynomial. On the other hand, by replacing the condition of the existence of the polynomial $e$ by other conditions, we prove that for a list of values indexed by a set of the form $\S(t)$ it is possible to construct a unique doubly periodic array $U$ containing the original values. Finally, we apply all those results to the locator decoding technique in abelian bivariate codes. Specifically, we prove that for any abelian code whose defining set contains a set of the form $\tau+\S(t)$, for some $\tau\in\N\times\N$, its correction capability is at least $t$ and in any transmision with no more than $t$-errors, it is possible to decode succesfully by the BMSa.

\section{Notation and preliminaries}

Let $\F$  be a finite field with $q$ elements, where $q$ is power of a prime number, let $r_i$ be positive integers, for $i\in \{ 1,2\}$, and $r=r_1\cdot r_2$.  We denote by $\Z_{r_i}$ the ring of integers modulo $r_i$. We always write its elements as canonical representatives. When necessary we write $\overline{a}\in\Z_k$ for any $a\in\Z$.  

For each $i\in \{ 1,2\}$, we denote by $R_{r_i}$ (resp. $\R_{r_i}$) the set of  $r_i$-th roots of unity (resp.  $r_i$-th primitive roots of unity) and define $R=R_{r_1}\times R_{r_2}$ ($\R=\R_{r_1}\times \R_{r_2}$). Throughout this paper, we fix $\Le|\F$ as a extension field containing $R_{r_i}$, for $i=1,2$.

We define the quotient algebra $\F(r_1,r_2)=\F[X_1,X_2]/\langle X_1^{r_1}-1, X_2^{r_2}-1\rangle $. Its elements are identified with polynomials. We denote by $I$ the set $\Z_{r_1}\times \Z_{r_2}$ and  we write the elements $f \in  \F(r_1,r_2)$ as $f=\sum a_m \bf{X}^m$, where $m=(m_1,m_2)\in I$ and $\bf{X}^m=X_1^{m_1}\cdot X_2^{m_2}$. The weight of any $f\in \F(r_1,r_2)$ is denoted by $\omega(f)$. Given a polynomial $f \in \F[\X]$, we denote by $\overline{f}$ its image under the canonical projection onto $\F(r_1,r_2)$, when necessary. For $f=f(X_1,X_2) \in \F[X_1,X_2]$ and $\bar{\alpha}=(\alpha_1,\alpha_2)\in R$, we write $f(\bar{\alpha})=f(\alpha_1,\alpha_2)$. For $m=(m_1,m_2)\in I$, we write $\bar{\alpha}^{m} = (\alpha_1^{m_1},\alpha_2^{m_2})$.

Throughout this paper we assume that $\gcd(r,q)=1$; so that, the algebra $\F(r_1, r_2)$ is semi simple. As a consequence, it is a known fact that every ideal $C$ in the algebra   $\F(r_1, r_2)$ is totally determined by its \textbf{root set} or \textbf{set of zeros}, namely
$$Z(C)=\left\{\bar{\alpha}\in  R \tq f(\bar{\alpha})=0,\;\; \mbox{ for all }\; f\in C \right\}.$$  For a fixed $\bar{\alpha}\in \R$, the ideal $C$ is  determined by its \textbf{defining set}, with respect to $\bar{\alpha}$, which is defined as 
$$\D_{\bar{\alpha}}(C) = \left\{ m\in I \tq \bar{\alpha}^{m}\in Z(C)\right\}.$$ 

It is easy to see that the notions of set of zeros and defining set may be considered for any set of either polynomials or ideals in $\F(r_1,r_2)$ (or $\Le(r_1,r_2)$); moreover, it is known that for any $G\subset \F(r_1,r_2)$ (or $\Le(r_1,r_2)$) and $\overline \alpha \in \R$ we have $D_{\overline{\alpha}}(G)=D_{\overline{\alpha}}(\langle G\rangle)$. In \cite{Blah,Cox}, the defining set is  considered for ideals $P$ in $\Le[\mathbf{X}]$. From the definition, it is clear that $D_{\overline{\alpha}}(P)=D_{\overline{\alpha}}(\overline{P})$, where $\overline{P}\in\Le[\mathbf{X}]$ is the canonical projection of $P$ onto $\Le(r_1,r_2)$.

\section{The Berlekamp-Massey-Sakata algorithm}

As we have commented, the BMSa is an iterative procedure (w.r.t. a total ordering) to construct a Groebner basis for the ideal of polynomials satisfying some linear recurring relations for a doubly periodic array. Let us set up basic terminology and some facts about it. We will also make slight modifications in order to improve its application.

We denote by $\N$ the set of natural numbers (including 0) and we define $\Sigma_0=\N\times \N$, as in \cite{Sakata 2}. We consider the partial ordering in $\Sigma_0$ given by $(n_1,n_2) \preceq (m_1,m_2) \Longleftrightarrow n_1\leq m_1\quad\text{and}\quad n_2\leq m_2.$ On the other hand, we will use a (total) monomial ordering \cite[Definition 2.2.1]{Cox}, denoted by ``$\leq_T$'', as in \cite[Section 2]{Sakata 2}. This ordering will be either the lexicographic order (with $X_1>X_2$) \cite[Definition 2.2.3]{Cox} or the (reverse) graded order (with $X_2>X_1$) \cite[Definition 2.2.6]{Cox}. Of course, any result in this paper may be obtained under the alternative lexicographic or graded orders. The meaning of ``$\leq_T$'' will be specified as required.

\begin{definition}\label{Sigmas y Delta rectangulo}
 For $s,k\in\Sigma_0$, we define
\begin{enumerate}
 \item $\Sigma_{s}=\left\{m\in \Sigma_0\tq s\preceq m\right\}$,
 \item  $\Sigma_{s}^k =\left\{m\in \Sigma_0\tq s\preceq m \;\;\text{and}\;\; m <_T k\right\}$ and 
 \item $\Delta_{s}=\left\{n\in \Sigma_0\tq n \preceq s\right\}$.
\end{enumerate}
\end{definition}

Given $m,n\in \Sigma_0$, we define $m+n$, $m-n$ (provided that $n\preceq m$)  and $n\cdot m$, coordinatewise, as it is usual. An infinite array or matrix is defined as $U=\left(u_n\right)_{n\in \Sigma_0}$; where the $u_n$ will always belong to the extension field $\Le$. In practice, we work with finite arrays. Due to some computations, we are interested to define them as infinite doubly periodic arrays (see \cite[p. 324]{Sakata 2}) and consider subarrays, as follows.

\begin{definition}\label{def arreglo doblemente periodico}
Let $U=\left(u_n\right)_{n\in \Sigma_0}$ be an infinite array.
 \begin{enumerate}
  \item We say that $U$ is a doubly periodic array of period $r_1\times r_2$ if the following property is satisfied: for $n=(n_1,n_2)$ and $m=(m_1,m_2)\in \Sigma_0$, we have that $n_i\equiv m_i\mod r_i$ for $i=1,2$ implies that $u_n=u_m$.
  
  \item If $U$ is a doubly periodic array of period $r_1\times r_2$, a finite subarray $u^l$ of $U$, with $l\in \Sigma_0$, is the array $u^l=\left(u_m\tq m\in \Sigma_0^l\cap \Delta_{(r_1-1,r_2-1)}\right)$. We denote it as $u^l \subset U$.
 \end{enumerate}
 In case $l>_T(r_1,r_2)$ then $u^l=\left(u_m\tq m\in \Delta_{(r_1-1,r_2-1)}\right)=\left(u_m\tq m\in I\right)$.
\end{definition}

Note that, in the case of period $r_1\times r_2$ we may identify $I=\Z_{r_1}\times\Z_{r_2}=\Delta_{(r_1-1,r_2-1)}$.

As it is well known, every monomial ordering is a well order, so that any $n\in \Sigma_0$ has a succesor. For the graded order we have
$$n+1=\begin{cases}
	    (n_1-1,n_2+1) & \text{if } n_1>0\\
            (n_2+1,0) & \text{if } n_1=0
      \end{cases}.$$

In the case of the lexicographic order, we have to introduce, besides the unique succesor with respect to the monomial ordering, another ``succesor'' (or next-step point) that we will only use for the recursion steps over $n\in \Delta_{(r_1-1,r_2-1)}$. We also denote it by $n+1$ as follows:
$$n+1=\begin{cases}
	    (n_1,n_2+1) & \text{if } n_2<r_2-1\\
            (n_1+1,0) & \text{if } n_2=r_2-1
      \end{cases}.$$
      
So, during the implementation of the BMSa (that is, results related with it), the succesor of $n\in \Delta_{(r_1-1,r_2-1)}$ will be donted by $n+1$, independently of the monomial ordering considered.

As in \cite[p. 323]{Sakata 2}, for any $f\in \Le[\bf{X}]$ or $f\in \Le(r_1,r_2)$, we denote the leading power product exponent of $f$, with respect to ``$\leq_T$'' by $LP(f)$. Of course $LP(f)\in \Sigma_0$. For a subset $F\subset \Le[\bf X]$, we denote $LP(F)=\{LP(f)\tq f\in F\}$.

\begin{definition}[see \cite{Sakata 2}]\label{def recurrencia lineal}
 Let $U$ be a doubly periodic array with entries in $\Le$, $f\in \Le[\bf X]$, $n\in \Sigma_0$ and $LP(f)=s$. We write $f=\sum_{m\in \supp(f)}f_m\bf{X}^m$ and define 
 \[f[U]_n=\begin{cases}
           \displaystyle{ \sum_{m\in\supp(f)}f_m u_{m+n-s}}& \text{if }  n\in\Sigma_s\\
           0 & \text{otherwise}
          \end{cases}.\]
The equality $f[U]_n=0$ will be called a \textbf{linear recurring relation}. 

If $f[U]_n=0$ for $n\in \Sigma_0$, we will say that the polynomial \textbf{$f$ is valid for $U$ at $n$}.
\end{definition}

Now we recall the following definitions in \cite{Sakata 2}.
\begin{definitions} Let $U$ doubly periodic and $f\in \Le[\bf X]$, with $LP(f)=s$.
 \begin{enumerate}
   \item We say that $f$ generates $U$ and write $f[U]=0$, if $f[U]_n=0$ at any $n\in \Sigma_0$.
   \item For any $u=u^k\subset U$, we say that $f$ generates $u$ if $f[U]_n=0$ at every $n\in \Sigma_s^k$ and we write $f[u]=f[u^k]=0$. In case $\Sigma_s^k=\emptyset$ we define $f[u]=0$.
    \item For any $u=u^k\subset U$, we say that $f$ generates $u$, up to $l<_T k$, if $f[u^l]=0$. 
    
 \item Let $u=u^k\subset U$.
 \begin{enumerate}
  \item  We write the set of generating polynomials for $u$ as
  \[\bL(u)=\{f\in \Le[\bf X]\tq f[u]=0\}.\]
\item  We write the set of generating polynomials for $U$ as
  $$\bL(U)=\left\{f\in\Le[\bf X] \tq f[U]=0\right\},$$
  which was originally called $VALPOL(U)$ \cite[p. 323]{Sakata 2}.
 \end{enumerate}

 \end{enumerate}
\end{definitions}

 \begin{remark}\label{hechos sobre lambda de U}
 By results in \cite{Blah}, \cite{Sakata 2} and \cite{Sakata} we have the following facts:
 
\begin{enumerate}
  \item $\bL(U)$ is an ideal of $\Le[\bf X]$.
  \item Setting $\overline{\bL(U)}=\left\{\overline{g}\mid g\in \bL(U)\right\}$, and seeing the elements of $\Le(r_1,r_2)$ as polynomials, we have that $\overline{\bL(U)}$ is an ideal, and  $\overline{\bL(U)}=\Le(r_1,r_2)\cap \bL(U)$.
 \end{enumerate}
\end{remark}

\subsection{ Delta sets and defining points.}\label{parrafo de conjuntos delta y puntos de def}

Let $0<d\in\N$ and consider the sequence $s^{(1)},\dots,s^{(d)}$ in $\Sigma_0$ satisfying
\begin{equation}\label{desigualdades de los puntos de def}
 s^{(1)}_1>\ldots>s^{(d)}_1=0 \quad\text{and}\quad 0=s^{(1)}_2<\ldots<s^{(d)}_2.
\end{equation}

Now we set, for $i\in \{1,\dots,d\}$,
\begin{eqnarray}\label{los delta para G} \nonumber
\Delta_i&=&\left\{m\in \Sigma_0\tq m\preceq \left(s^{(i)}_1-1,s^{(i+1)}_2-1\right)\right\}_{1\leq i\leq d-1}\\
&=&\Delta_{(s^{(i)}_1-1,s^{(i+1)}_2-1)}
\end{eqnarray}
 and define  $\Delta=\bigcup_{i=1}^{d-1}\Delta_i$, which is called a \textbf{$\Delta$-set} (or delta-set), and the elements $s^{(1)},\dots,s^{(d)}$ are called its \textbf{defining points}.

We denote by $\fF$ the collection of sets $F=\left\{f^{(1)},\dots,f^{(d)}\right\}\subset \L[\bf X]$ such that $\{LP(f^{(i)})=s^{(i)}\tq i=1,\dots, d\}$ satisfy the condition of Inequalities~\eqref{desigualdades de los puntos de def}. We shall say that the elements $F\in\fF$ are of type $\Delta$ and we denote by $\Delta(F)$ the $\Delta$-sets determined by them.

 \begin{definition}\label{base de u}
 Let $U$ be doubly periodic and $u=u^k\subset U$.
 In the situation of paragraph above, we say that the set $F=\left\{f^{(1)},\dots,f^{(d)}\right\}$ is a minimal set of polynomials for $u$ if:
  \begin{enumerate}
   \item $F\subset \bL(u)$.
   \item $F\in \fF$; that is $\Delta(F)$ exists.
   \item If $0\neq g\in\L[\bf X]$ verifies $LP(g)\in \Delta(F)$ then $g\not\in\bL(u)$ (i.e. $g[u]\neq 0$).
  \end{enumerate}
 \end{definition}
 
 We denote by $\fF(u)$ the collection of the minimal sets of $u$.

 From \cite{Blah} and \cite{Sakata 2} we have the following properties of $\Delta$-sets and minimal sets of polynomials.
 
 \begin{remark}\label{los delta conjuntos y los minimales}Let $U$ be a doubly periodic array.
 \begin{enumerate}
 \item For any $l\in\Sigma_0$, $u=u^l\subset U$ and $F,F'\in \fF(u^l)$ we have that $\Delta(F)=\Delta(F')$, so that we may write $\Delta(u^l)$.
  \item $\Delta(u^l)\subseteq \Delta(U)$ for all $l\in \Sigma_0$ and if $k<_T l\in\Sigma_0$  then $\Delta(u^k)\subseteq \Delta(u^l)$.
  \item For any $l\in\Sigma_0$, the set $\Delta(u^l)$ always exists.
  \item The set $\Delta(U)$ is exactly the footprint (see \cite[Definition, p. 1615]{Blah}) of $\bL(U)$, and it is completely determined by any of its Groebner basis.
  \item For any $F\in \fF(u^l)$ we have that $F\subset \bL(U)$ implies $\langle F\rangle =\bL(U)$. In fact, $F$ is a Groebner basis for $\bL(U)$ by Definition~\ref{base de u} (3)  and \cite[Definition 2.5]{Cox}.
  \item As we have commented, for any $\overline{\alpha}\in\R$, the equality $\D_{\bar{\alpha}}\left(\overline{\bL(U)}\right)=\D_{\bar{\alpha}}\left(\bL(U)\right)$ holds. Then by \cite[Proposition 5.3.1]{Cox} or \cite[p. 1617, Theorem]{Blah} we have that $\left|\D_{\bar{\alpha}}\left(\overline{\bL(U)}\right)\right|=|\Delta(U)|$  (see also \cite[p. 1202]{Sakata}).
 \end{enumerate}
 \end{remark}

As we have mentioned, minimal sets for the ideal $\bL(U)$ are Groebner basis. In the case of finite arrays $u\subset U$, the set $\bL(u)$ is not an ideal; however, sometimes (see Example~\ref{ejemplo del problema de evaluar} and Proposition~\ref{forma normal en Lambda}) we will require minimal sets of $\bL(u)$ to be written in ``normal form'', as in \cite[p. 83]{Cox}. To do this, we need the following technical results.

\begin{lemma}\label{suma recurrencias grado diferente}
  Let $U$ be a doubly periodic array, $l\in \Sigma_0$, $u=u^l\subset U$ and $f,g\in \Le[\bf X]$, with  $LP(f)=s_f$ and $LP(g)=s_g$. If $s_f<_T s_g$ and $n\in \Sigma_{s_g}^l\neq \emptyset$ then $n-s_g+s_f\in \Sigma_{s_f}^l$ and $(f+g)[u]_n=f[u]_{(n-s_g+s_f)}+g[u]_n$.
\end{lemma}
\begin{proof}
 We write $f=\sum_{m\in \supp(f)}f_m\bf{X}^m$ and $g=\sum_{m\in \supp(g)}g_m\bf{X}^m$. If $s_f<_T s_g$ then, for each term we have $(f+g)_m u_{m+n-s_g}=f_m u_{m+(n-s_g+s_f)-s_f}+g_m u_{m+(n-s_g)}$. Now, $s_f<_T s_g$ implies that $n+s_f<_T n+s_g$, then $n+s_f-s_g<_T n<_T l$ and $s_f \preceq n+s_f-s_g$, clearly; so that $n-s_g+s_f\in \Sigma_{s_f}^l$ and $(f+g)[u]_n=f[u]_{(n-s_g+s_f)}+g[u]_n$.
\end{proof}

\begin{lemma}\label{Xmultiplos de pol}
 Let $U$ be a doubly periodic array, $l,t\in\Sigma_0$ and $u=u^l\subset U$. If $f\in \Lambda(u)$ then $\X^t f\in \Lambda(u)$.
\end{lemma}
\begin{proof}
 Set $s=LP(f)$. By hypothesis, for any $n\in \Sigma_s^l$, we have $0=\sum_{m\in \sup(f)}f_m u_{m+n-s}$. Now, if $t+s\geq_T l$ then $\X^tf\in \Lambda(u)$, by Definition~\ref{def recurrencia lineal}. Otherwise, for any $n\in \Sigma_{s+t}^l$, we have $\X^tf[u]_n=\sum_{m\in \sup(f)}f_m u_{(m+t)+n-(s+t)}=\sum_{m\in \sup(f)}f_m u_{m+n-s}=0$.
\end{proof}

As a direct consequence we have.

\begin{proposition}\label{combinaciones lineales en lambdas}
 Let $U$ be a doubly periodic array, $l,t\in\Sigma_0$ and $u=u^l\subset U$. If $f,g\in \Lambda(u)$ and $LP(f)+t\neq LP(g)$ then $\X^t f+g\in \Lambda(u)$
\end{proposition}
\begin{proof}
 Set $LP(f)=s_f$ and $LP(g)=s_g$. Suppose, first, that $s_f+t<_Ts_g$; then $LP(\X^tf+g)=s_g$. In case $s_g\geq_T l$ we have, by Definition~\ref{def recurrencia lineal}, that $\X^tf+g\in \Lambda(u)$. Otherwise, for any $n\in \Sigma_{s_g}^l$, by Lemma~\ref{suma recurrencias grado diferente}, $(\X^tf+g)[u]_n=\X^tf[u]_{(n-s_g+s_f+t)}+g[u]_n$, and $n-s_g+s_f+t\in\Sigma_{s_f+t}^l$. By Lemma~\ref{Xmultiplos de pol}, $\X^tf\in \Lambda(u)$, so that $\X^tf[u]_{(n-s_g+s_f+t)}=0$; hence $(\X^tf+g)[u]_n=0$. The other case is analogous.
\end{proof}

 Let $U$ be a doubly periodic array. Each iteration in the BMSa is a procedure to obtain a minimal set of polynomials for $u=u^{l+1}\subset U$ from such a set $u^l$ and the $\Delta$-set $\Delta(u^l)$. The procedure works as follows. Suppose it has been constructed a minimal set of polynomials $F$ for $\bL(u^l)$ and define $F_N=\{f\in F\tq f[u]_l\neq 0\}$. Then one has to replace, one by one, each element $f\in F_N$ by one or more polynomials $f'\in \bL(u^{l+1})$ until get a new minimal set of polynomials $F'$.
 
 To ensure that the procedure will be successful, Sakata's paper \cite{Sakata 2} (see also \cite{Blah}) provided results in two directions. The first one is finded in the following theorem, that says us that once one has produced sufficiently many terms, there is only one way to get subsequent terms.
 
 \begin{theorem}[Agreement Theorem \cite{Blah}]\label{teo coincidencia 2 var}
Let $f,g\in\L[\X]$ such that $LP(f)=s$ and $LP(g)=t$, and let $u^l\subset u$. Suppose that $f[u^l]=g[u^l]=0$. If $s+t\preceq l$ then $g[u]_l=f[u]_l$.
 \end{theorem}
 
 The first application of the Agreement Theorem refers to the increase of $\Delta$-sets.
 
 \begin{corollary}[Sakata-Massey Theorem \cite{Blah}]\label{teo Sakata-Massey}
 Let $f\in \L[\X]$ such that $LP(f)=s$ and $f[u^l]=0$, but $f[u]_l\neq 0$. If $g\in\L[\X]$ is such that $LP(g)=l-s$ then $g\not\in\bL(u^{l+1})$. 
 
 As a consequence, if $f\in\bL(u^l)$, but $f[u]_l\neq 0$ then $l-s\in \Delta(u^{l+1})$.
\end{corollary}

The second group of results, provided by Sakata, refers directly to the construction of replacement polynomials, and it also needs certain produced terms.

\begin{lemma}(\cite[Lemma 5]{Sakata 2}). Let $k<_Tl\in \Sigma_0$, and $f,g\in\L[\X]$ with $LP(f)=s$ and $LP(g)=t$. Suppose that $g\in \bL(u^k)$, but $g[u]_k=v\neq 0$ and $f\in \bL(u^l)$, but $f[u]_l=w\neq 0$.
 We define 
 \begin{eqnarray*}
  r_1&=&\max\{s_1,l_1+t_1-k_1\},\\
  r_2&=&\max\{s_2,l_2+t_2-k_2\}\;\text{and}\\
  \mathbf{e}&=&r-l+k-t.
 \end{eqnarray*}
Then, setting $r=(r_1,r_2)$,
 $$ h_{(f,l,s;g,k,t)}=\bf{X}^{r-s}f-\frac{w}{v}\bf{X}^{\mathbf{e}}g^{(b)}\in\bL(u^{l+1}).$$
\end{lemma}

The reader may see that the form of the polynomial above is the reason of the construction of the set described in the following remark.

 \begin{remark}\cite[p. 327]{Sakata 2}\label{conjunto G}. Let $U$ be doubly periodic and $u=u^l\subset U$. For any minimal set of polynomials $F=\left\{f^{(1)},\dots,f^{(d)}\right\}\in \fF(u^l)$  one may see that there exists another set of polynomials $G=\{g^{(1)},\dots,g^{(d-1)}\}$ such that, for each $i\in \{1,\dots,d-1\}$
 \begin{enumerate}
  \item $g^{(i)}$ determines the set $\Delta_i$ in Equalities~\eqref{los delta para G}.
  \item There is $k^{(i)}<_Tl$ such that $g^{(i)}\in \bL(u^{k^{(i)}})$ but $g^{(i)}[u]_{k^{(i)}}\neq 0$.
 \end{enumerate}
 \end{remark}
 
The set $G$ above is called an auxiliary set of polynomials associated to $F$. 

Finally, we construct replacement polynomials as a consequence of the lemma above.
 
 \begin{proposition}[Berlekamp Procedure] (\cite[Lemma 6]{Sakata 2}): Supposse we have constructed a minimal set $F=\{f^{(1)},\dots,f^{(d)}\}\in \fF(u^l)$ and an auxiliary set (see Remark~\ref{conjunto G}) $G=\{g^{(1)},\dots,g^{(d-1)}\}$. Set $LP(f^{(i)})=s^{(i)}$, for $i=1,\dots,d$.

Let $f^{(a)}\in F$ and $g^{(b)}\in G$ such that $f^{(a)}\in \bL(u^l)$, $g^{(b)}\in \bL(u^{k})$, for some $k<_T l\in I$, with $f^{(a)}[u]_l=w_a\neq 0$ and $g^{(b)}[u]_{k}=v_b\neq 0$. 
We define
 \begin{eqnarray*}
  r_1&=&\max\{s_1^{(a)},l_1-s_1^{(b)}+1\},\\
  r_2&=&\max\{s_2^{(a)},l_2-s_2^{(b+1)}+1\}\;\text{and}\\
  \mathbf{e}&=&\left(r_1-l_1+s_1^{(b)}-1,\,r_2-l_2+s_2^{(b+1)}-1 \right).
 \end{eqnarray*}
Setting $r=(r_1,r_2)$ we have that
 $$ h_{f^{(a)},g^{(b)}}=\bf{X}^{r-s^{(a)}}f^{(a)}-\frac{w_a}{v_b}\bf{X}^{\mathbf{e}}g^{(b)}\in\bL(u^{l+1}).$$
\end{proposition}

We note that $s_1^{(b)}$ and $s_2^{(b+1)}$ refers to elements of $F$ and not $G$. \\

Let us keep the notation of the proposition above, with $F_N=\{f\in F\tq f[u]_l\neq 0\}$. As one may see, when some $f^{(i)}\in F_N$ is going to be replaced one has to choose those elements $g^{(j)}\in G$ that will be used in order to apply the Berlekamp Procedure. These choices depend on the possible new defining points determined by the position of the point $l-s^{(i)}$ with respect to $\Delta(u^{l})$. Let us comment it briefly. If $l-s^{(i)}\in \Delta(u^l)$ then one may see that $s^{(i)}$ will be again a defining point, given us the defining points of Type 1, below. In the case $l-s^{(i)}\not \in \Delta(u^l)$ one has to consider the greatest $j\in \{1,\dots,d\}$ such that $s^{(j)}\preceq l-s^{(i)}$, so that, by the Agreement Theorem, $f^{(j)}\in F_N$, too. Then, one has to consider jointly the position of the points (if $i\neq j$)  $l-s^{(i)}$ and $l-s^{(j)}$ with respect to $\Delta(u^{l})$, obtaining the rest of the following classification.

\begin{theorem}[Classification of defining points]\label{tipos de m}
~
 \begin{description}
  \item[] From  \cite[Lemma 7 and Theorem 1]{Sakata 2} 
  \item[Type 1.]~ $S=s^{(i)}$, where $f^{(i)}\in F_N$ and $l\in s^{(i)}+\Delta(u^l)$.
  \item[] From \cite[Lemma 7, Remark 2 (p. 329) and Lemma 8]{Sakata 2}
 \item[Type 2.]~ $S=\left(l_1-s_1^{(i)}+1,l_2-s_2^{(i+1)}+1\right)$; where $f^{(i)},f^{(i+1)}\in F_N$, with $1\leq i<d$.
 \item[] The last two types comming from  \cite[Lemma 7, Remark 2 (p. 329) and Lemma 9]{Sakata 2}.
\item[Type 3.]~ $S=\left(l_1-s_1^{(j)}+1,s_2^{(i)}\right)$;  with $j<d$, $f^{(i)},\,f^{(j)}\in F_N$ and $s^{(i)}\prec S$; moreover, $l\succeq s^{(i)}+s^{(j)}$ and for all $k>i$ we have $l_2-s_2^{(k)}<s_2^{(j)}$.      
\item[Type 4.]~ $S=\left(s_1^{(i)},l_2-s_2^{(j)}+1\right)$, where $f^{(i)},\,f^{(j)}\in F_N$ and $s^{(i)}\prec S$, with $2\leq j$; moreover, $l\succeq s^{(i)}+s^{(j)}$ and for all $k<j$ we have $l_1-s_1^{(k)}<s_1^{(i)}$.
\end{description} 
\end{theorem}

\subsection{The Berlekamp-Massey-Sakata algorithm}\label{BMSa}
 
\begin{procedure}\cite[Theorem 1]{Sakata 2}.\label{proc 1}
 If $f^{(i)}\in F_N$ and $l\in s^{(i)}+\Delta(u^l)$.
\begin{enumerate}
 \item Find $1\leq j\leq d-1$ such that $l_1<s_1^{(i)}+ s_1^{(j)}$ and    $l_2<s_2^{(i)}+ s_2^{(j+1)}$.
 \item In the set F we replace $f^{(i)}$ by $h_{f^{(i)},g^{(j)}}$ (obtained by the Berlekamp procedure). The point $s^{(i)}$ will be a defining point of $\Delta(u^{l+1})$ as well.
\end{enumerate}
\end{procedure}

\begin{procedure}\cite[Theorem 2]{Sakata 2}.\label{proc 2}  If $f^{(i)}\in F_N$ and $l\not\in s^{(i)}+\Delta(u^l)$ then one considers all the following defining points and constructions $h_{f^{(a)},g^{(b)}}$ to replace $f^{(i)}$ (and, possibly, some elements of $G$) with the suitable new polynomials in order to get a new $F\in\fF(u^{l+1})$.

\begin{enumerate}
 \item $S=\left(l_1-s_1^{(i)}+1,l_2-s_2^{(i+1)}+1\right)$; with $f^{(i+1)}\in F_N$ and $1\leq i<d$. 
 Then find $k\in\{1,\dots,d\}$ such that $s^{(k)}\prec S$ and set  $h_{f^{(k)},g^{(i)}}$.

\item $S=\left(l_1-s_1^{(k)}+1,s_2^{(i)}\right)$; for some $k<d$, with $f^{(k)}\in F_N$ and $s^{(i)}\prec S$. Then set $h_{f^{(i)},g^{(k)}}$.

\item $S=\left(l_1+1,s_2^{(i)}\right)$; (the case $k = d$). 
Then set $h=X_1^{l_1-s_1^{(i)}+1}\cdot f^{(i)}$.

\item $S=\left(s_1^{(i)},l_2-s_2^{(j)}+1\right)$ for $j\geq 2$ with $f^{(j)}\in F_N$ and $s^{(i)}\prec S$. Then set $h_{f^{(i)},g^{(j-1)}}$.

\item $S=\left(s_1^{(i)},l_2+1\right)$; (the case $j = 1$). Then set $h=X_2^{l_2-s_2^{(i)}+1}\cdot f^{(i)}$.
\end{enumerate} 
\end{procedure}

Now, we can show a brief scheme of the Sakata's algorithm. See \cite[p. 331]{Sakata 2} for a detailed description.

\begin{algorithm*}[Sakata]
 We start from a finite doubly periodic array, $u\subset U$.
 \begin{enumerate}
  \item[$\bullet$] Initialize $F:=\{1\}$, $G:=\emptyset$ and $\Delta:=\emptyset$.
 \item[$\bullet$] For $l\geq (0,0)$,
 \item For each $f^{(i)}\in F$ for which $f^{(i)}\in F_N$ we do
 \begin{description}
  \item[-] If $l\in s^{(i)}+\Delta(u^l)$ then replace $f^{(i)}$ by Procedure~\ref{proc 1}. 
  \item[-] Otherwise, replace $f^{(i)}$ by one or more polynomials by Procedure~\ref{proc 2}.
 \end{description}
 \item Then form the new $F'$ selecting from among $F\setminus F_N$ and those polynomials constructed by the two procedures; $G'$ selecting from among $G\cup F_N$, and $\Delta(u^{l+1})$.
 \item Set $l:=l+1$.
 \end{enumerate}
\end{algorithm*}

For the sake of a better general view of the algorithm, we want to comment briefly the fact that the set $G'$ may be obtained from $G\cup F_N$. For any $f^{(i)}\in F_N$ and $l-s^{(i)}\not\in \Delta(u^l)$ the set $\Delta_{l-s^{(i)}}$ has a nice property: if there exists $a,b\in \Sigma_{0}^{l+1}$, with $b\preceq a$ such that $\Delta_{l-s^{(i)}}\subseteq \Delta_{a-b}\subset \Delta(u^{l+1})$ then $a=l$ and $b=s^{(i)}$; so that the set $\Delta_{l-s^{(i)}}$ must be one of those sets $\Delta_k$ considered in Remark~\ref{conjunto G}(1) and hence $f^{(i)}$ may be considered to be an element of $G'$.
 
The following remark summarizes the construction of $\Delta$-sets and defining points.
 
 \begin{remark}\label{resumen condiciones crece delta}
 Suppose that $f\in F\in \fF(u^l)$, but $f[u]_l\neq 0$. Then one of the following cases must happen:
 \begin{enumerate}
  \item  $l-LP(f)\in\Delta(u^l)$ and hence $LP(f)$ will be again a defining point of $\Delta(u^{l+1})$ by applying Procedure~\ref{proc 1}.
  
 \item $l-LP(f)\not\in\Delta(u^l)$ and hence, $\Delta(u^{l+1})$ will have at least one point more, $l-LP(f)$ itself. In this case,  there exists (one or more) replacement polynomials by applying Procedure~\ref{proc 2}.
  \end{enumerate}
\end{remark}

Once one has used all known suitable values of the given array; say $u^l\subset U$, and one has constructed the last minimal set of polynomials, say $F$, one proceeds to reduce it to its normal form (see Proposition~\ref{forma normal en Lambda}, below) to find the desired Groebner basis for $\bL(U)$. However, the normal form is not always a Groebner basis for $\bL(U)$. In \cite[Section 6]{Sakata 2} some conditions are given to ensure the existence of a Groebner basis, $F'$ that has a uniqueness property w.r.t. some system of equations. From $F'$ we may construct a doubly periodic array $U'$ such that $u'^{l}=u^{l}$, but we cannot be sure that neither $\langle F'\rangle=\langle F\rangle$ nor $\langle F'\rangle=\bL(U)$.  To find sufficient conditions to ensure that the reduced set (to its normal form) is the desired Groebner basis is an  interesting problem as we have commented in the introduction. In \cite{Cox et al Using,Hackl,rubio} and \cite{Sakata 2} one may find some termination criteria based on the computation of bounds that depend on the analysis of the shape of all possible sets $\Delta(U)$.

In Section~\ref{seccion de condiciones suficuentes}, we will study the problem of finding  a set of indexes, that we call $\S(t)$ (see Definition~\ref{def conjunto adecuado}), such that, along with some conditions, it guarantees that the set of polynomials obtained at the last iteration in the BMSa over such a set is exactly a Groebner basis for $\bL(U)$ (theorems~\ref{condicion suficiente} and \ref{condicion suficiente no locator}).

\section{A new framework for locator ideals}\label{el ideal locator}

The locator decoding method was originally introduced in the context of decoding error correcting codes; however, as it happens with the Berlekamp-Massey algorithm in the case of BCH codes, it may be described in a more general framework.

Throughout this section we fix an element $\overline{\alpha}\in \R$.  Let $e\in \F(r_1,r_2)$ and let $S=\left(s_n\right)_{n\in \Sigma_0}$ be the doubly periodic array defined by the syndromes $s_n=e\left(\overline{\alpha}^n\right)$. We want to find the coefficients of $e$ by using some (less as possible) values of $S$ (almost all of them, unknown in practice). So the problem is to determine a (minimal) set of indexes, $\A\subset \Z_{r_1}\times \Z_{r_2}$, for which, the application of the BMSa  solely on the values of the set $\{s_n\tq n\in \A\}$ gives us a Groebner basis for $\bL(U)$ (see Definition~\ref{def de U}), and then we may get the polynomial $e$ and all the values of $S$.

To achieve the solution we will need to impose some restrictions on the set $\A$. The first one says that we will need it to be contained in a rectangle of indexes of (known values of) $S$; that is, there must exist an element $\tau\in \Z_{r_1}\times \Z_{r_2}$ such that $\tau\in \A$ and $\tau \preceq n$ for all $n\in \A$. Now we introduce the notion of locator ideal. We note that, the classical notion in \cite{Blah,Sakata 2} defines the error-locator ideal, $I_e$ as an ideal of the polynomial ring $\Le[\mathbf{X}]$; however, we prefer work in the context of the ring $\Le(r_1,r_2)$.
 \begin{definition}
  Let $e$ be a polynomial in $\F(r_1,r_2)$. The locator ideal of $e$ is
  \[L(e)=\left\{f\in \Le(r_1,r_2)\tq f(\bar{\alpha}^n)=0,\;\forall n\in \supp(e)\right\}.\]
 \end{definition}
 
Having in mind that $\Le|\F$ is a splitting field, it is easy to see that $\D_{\bar{\alpha}}\left(L(e)\right)=\supp(e)$. Therefore, our objective is to find  the defining set of $L(e)$ and hence $\supp(e)$. Then we solve a system of equations to get the coefficients of $e$ (in case $q>2$). To find the defining set of $L(e)$, we shall connect it to linear recurring relations as follows. Based on the mentioned syndromes of the received polynomial, we are going to determine a suitable doubly periodic array $U=\left(u_n\right)_{n\in \Sigma_0}$ such that $L(e)=\overline{\bL(U)}$ (see Remark~\ref{hechos sobre lambda de U}). We first consider (theoretically) the following slightly modified table of syndrome values of $e\in \F(r_1,r_2)$.

\begin{definition}\label{def de U}
 Let $e\in \F(r_1,r_2)$, $\tau\in \Z_{r_1}\times \Z_{r_2}=I$ and define $U=\left(u_n\right)_{n\in \Sigma_0}$ such that $u_n= e\left(\bar{\alpha}^{\tau+n}\right)$. We call $U$ the syndrome table afforded by $e$ and $\tau$.
\end{definition}

Note that $U$ is an infinite doubly periodic array.The proof of the following theorem is (\textit{mutatis mutandi}) similar to that of \cite[p. 1202]{Sakata}.

\begin{theorem}\label{igualdad ideales}
 Let $U$ be the syndrome table afforded by $e$ and $\tau$. For any $f\in \Le(r_1,r_2)$ the following conditions are equivalent:
 \begin{enumerate}
  \item $f\in L(e)$.
  \item $\sum_{s\in\supp(e)}e_{s}\bar\alpha^{s\cdot n} f\left(\bar\alpha^{s}\right)=0,\;\text{for all}\; n\in \Sigma_\tau$.
  \item $f\in \overline{\bL(U)}$.
 \end{enumerate}
Consequently, $L(e)=\overline{\bL(U)}$.
\end{theorem}
\begin{proof}
$[1)\Rightarrow 2)]$.  
 Suppose that $f\left(\overline \alpha^s\right)=0$ for all $s\in \supp(e)$. Clearly, $\sum_{s\in\supp(e)}e_{s}\overline \alpha^{s\cdot n} f\left(\overline\alpha^s\right)=0$, for all $n\in \Sigma_\tau$ and we are done.
 
 $[2)\Rightarrow 1)]$. Conversely, assume \textit{(2)}. Set $e'_{s}=e_{s} f\left(\overline\alpha^s\right)$ and $e'(\X)=\sum_{s\in\supp(e)}e'_{s}\X^s$. By hypothesis, $e'\left(\overline\alpha^{n}\right)=0$, for all $n\in \Sigma_\tau$. Set $\bar\alpha=(\alpha_1,\alpha_2)$ and $\tau=(\tau_1,\tau_2)$. We write $e'(X_1,X_2)=\sum_{i=0}^{r_1-1}e'_i(X_2)X_1^i$ and we consider, for each $k\in\Z_{r_2}$, the one variable polynomial $P=e'(X_1,\alpha_2^{k+\tau_2})=\sum_{i=0}^{r_1-1}e'_i(\alpha_2^{k+\tau_2})X_1^i$. We know that $P$ has distinct roots $\alpha_1^{\tau_1},\dots,\alpha_1^{r_1-1+\tau_1}$ because they are abscissa of $\Sigma_{\tau}$; that is, $e'\left(\alpha_1^{\tau_1},\alpha_2^{k+\tau_2}\right)=e'\left(\overline\alpha^{\tau+(0,k)}\right)=0$, because $\tau+(0,k)\in \Sigma_\tau$. So, we have more roots than the degree of $P$ and then $P=0$, which means that $e'_i(\alpha_2^{k+\tau_2})=0$, for each $k=0,\dots,r_2-1$. This, in turn, implies that there are more distinct roots than the degree of $e'_i(X_2)$; so that $e'_i(X_2)=0$. Hence, $e'(X_1,X_2)=0$. In particular, for $s\in\supp(e)$ we have $e_{s}f(\overline\alpha^s)=0$, but  $e_{s}\neq 0$, so that $f(\overline\alpha^s)=0$.
 
 $[2) \Leftrightarrow 3)]$. Let $f\in \L(r_1,r_2)$, with $l=LP(f)$. The following equalities, for all  $n\in\Sigma_l$ will give us the desired result. Setting $t=n-l$ we have that
\begin{eqnarray*}
 \sum_{k\in\supp(f)} f_k u_{k+n-l}&=& \sum_{k\in\supp(f)} f_k u_{t+k}=\\&=& \sum_{k\in\supp(f)} f_{k} e\left(\bar\alpha^{t+k+\tau}\right)=\\
 &=&\sum_{k\in\supp(f)} f_{k}\sum_{s\in\supp(e)}e_{s}\bar\alpha^{s(t+k+\tau)}= \\&=& \sum_{k\in\supp(f)}f_{k}\sum_{s\in\supp(e)}e_{s}\bar\alpha^{s(t+\tau)}\bar\alpha^{sk}= \\
 &=&\sum_{s\in\supp(e)}e_{s}\bar\alpha^{s(t+\tau)} f\left(\bar\alpha^s\right). 
\end{eqnarray*}

So that,

\begin{equation*}
\sum_{k\in\supp(f)} f_k u_{k+n-l}=\sum_{s\in\supp(e)}e_{s}\bar\alpha^{s(t+\tau)} f\left(\bar\alpha^s\right).
\end{equation*}

Now, $f\in\overline{\bL(U)}$ if and only if, for all $n\in\Sigma_l$ we have that $0=\sum_{k\in\supp(f)} f_k u_{k+n-l}$. This is equivalent to say that $\sum_{s\in \supp(e)}e_s\overline{\alpha}^{s(t+\tau)}f\left(\bar\alpha^s\right)=0$ for all $t\in \Sigma_0$.
\end{proof}

Theorem \ref{igualdad ideales} says that if $F$ is a Groebner basis of $\bL(U)$, then 
\begin{equation}\label{igualdad dominios de definicion}
 \D_{\bar{\alpha}}(L(e))=\D_{\bar{\alpha}}(\overline{\bL(U)})=\D_{\bar{\alpha}}(F)
\end{equation}
 according to the notation of Section II.

 \section{Sufficient conditions to find Groebner basis for $\bL(U)$}\label{seccion de condiciones suficuentes}

Throughout this section we fix an element $\overline{\alpha}\in \R$. According to our stated problem, we want to find a set $\A$ in order  to obtain a Groebner basis for $\bL(U)$ by the BMSa. The first results are given in terms of the weight of the polynomial $e$. We recall that $\omega(e)\leq t$ implies $|\Delta(U)|\leq t$ by Equality~\eqref{igualdad dominios de definicion} and Remark~\ref{los delta conjuntos y los minimales}.6.

\begin{lemma}\label{cota para l2}
 Let $U$ be the syndrome table afforded by $e$ and $\tau$, with $\omega(e)\leq t\leq \lfloor\frac{r_i}{2}\rfloor$, with $i=1,2$, as in Definition~\ref{def de U}. Suppose that, following the BMSa we have constructed, for $l=(l_1,l_2)$, and $u=u^l$, the sets  $\Delta=\Delta(u)$ and $F\in\fF(u)$.  We also suppose that there is $f\in F$ such that $f[u]_l\neq 0$ and $l\not\in LP(f)+\Delta$; that is, the delta-set will increase, by Corollary~\ref{teo Sakata-Massey} (see Remark~\ref{resumen condiciones crece delta}).
 Then, the following inequality holds:
 $$l_1+l_2\leq t.$$
\end{lemma}
\begin{proof} Following the notation in Paragraph~\ref{parrafo de conjuntos delta y puntos de def}, we set $F=\left\{f^{(1)},\dots,f^{(d)}\right\}$, $f=f^{(i)}$ and $LP(f)=s^{(i)}$. We also have that $l-s^{(i)}\in \Delta(u^{l+1})$; in fact, $\Delta_{l-s^{(i)}}\subset \Delta(u^{l+1})$.
\\

\textbf{Case 1.} Suppose that $l_1\geq s^{(1)}_1$.\\

\textbf{Case 1a.} Suppose that $i=d$.

Then, we have that  $|\Delta|\geq s^{(d)}_2+s^{(2)}_2(s^{(1)}_1-1)$. Note that if $d=2$ and $s^{(1)}_1=1$, the second summand would have no contribution. From here, we see that it is possible to add at most $t-s^{(d)}_2-s^{(2)}_2(s^{(1)}_1-1)$ points to $\Delta$. Now we have $l-s^{(d)}=(l_1,l_2-s^{(d)}_2)$, and so, it has been added at least $(l_2-s^{(d)}_2+1)(l_1-s^{(1)}_1+1)$ points. So,
\begin{eqnarray*}
 0&\leq& t-s^{(d)}_2-s^{(2)}_2(s^{(1)}_1-1)-(l_2-s^{(d)}_2+1)(l_1-s^{(1)}_1+1),\\
 0&\leq& t-s^{(d)}_2-s^{(2)}_2(s^{(1)}_1-1)-\\&&-(l_2-s^{(d)}_2+1)(l_1-s^{(1)}_1) -l_2+s^{(d)}_2-1,\\
 l_2&\leq& t-s^{(2)}_2(s^{(1)}_1-1)-(l_2-s^{(d)}_2+1)(l_1-s^{(1)}_1)-1,\\
 l_2&\leq& t-s^{(1)}_1+1-l_1+s^{(1)}_1-1.
\end{eqnarray*}
because $s^{(2)}_2>0$ and also $l_2-s^{(d)}_2+1>0$, and from here
$l_2\leq t-l_1$.\\

\textbf{Case 1b.} Suppose that $i=1$.

In this case $|\Delta|\geq s^{(d)}_2+s^{(2)}_2(s^{(1)}_1-1)$. As above, from here we see that the number of points which is possible to add is at most $t-s^{(d)}_2-s^{(2)}_2(s^{(1)}_1-1)$.  Unlike the previous case, here we will have to divide into cases, again.\\

\textbf{Case 1b(i).} Suppose that $l_1-s^{(1)}_1\geq s^{(1)}_1$; that is ($l_1> 2s^{(1)}_1-1$). Then $l-s^{(i)}$, belongs to the vertical line $(l_1-s^{(1)},0),\dots,(l_1-s^{(1)},r_2-1)$ out of $\Delta$. Then we add $(l_2+1)(l_1-s^{(1)}_1+1)$ points. So, 

\begin{eqnarray*}
 0&\leq& t-s^{(d)}_2-s^{(2)}_2 (s^{(1)}_1-1)-(l_2+1)(l_1-s^{(1)}_1+1)\\
 0&\leq& t- s^{(d)}_2-s^{(2)}_2 (s^{(1)}_1-1)-(l_2+1)-\\&&-(l_2+1)(l_1-s^{(1)}_1)\\
  &&\text{as  }l_2+1\geq 1\\
 0&\leq& t- s^{(d)}_2-s^{(2)}_2 (s^{(1)}_1-1)-l_2-1-l_1+s^{(1)}_1 \\
 l_2&\leq&t- s^{(d)}_2-(s^{(2)}_2-1) (s^{(1)}_1-1)-l_1\\
  l_2&\leq&t-l_1
\end{eqnarray*}

\textbf{Case 1b(ii).} Suppose that $l_1-s^{(1)}_1< s^{(1)}_1$. Then by hypothesis and the definition of $\Delta$-set, we know that there exists $j\in\{1,\dots,d\}$, such that $s^{(j)}\preceq l-s^{(1)}$, and then $f^{(j)}\in F_N$ with $j\neq 1$. If $j=d$, we are done. Otherwise, we leave $f^{(i)}$ and consider $f^{(j)}$. Now we have $s^{(1)}\preceq l-s^{(j)}$, and then $s^{(1)}_1\leq l_1-s^{(j)}_1$. This new situation will be considered in Case 1c(i), bellow.\\

\textbf{Case 1c.} Suppose that $1<i<d$.

Here, we have the bound
\[|\Delta|\geq s^{(d)}_2+s^{(i)}_2 s^{(i)}_1+s^{(2)}_2(s^{(1)}_1-s^{(i)}_1-1)\]
and we divide again in cases.\\

\textbf{Case 1c(i).} Suppose that $l_1-s^{(i)}_1\geq s^{(1)}_1$; that is, ($l_1\geq s^{(i)}_1+s^{(1)}_1$), we are in a vertical line (as above) out of $\Delta$. Then we add at least $(l_2-s^{(i)}_2+1)(l_1-s^{(i)}_1-s^{(1)}_1+1)$ points. So,
\begin{eqnarray*}
 0&\leq& t-s^{(d)}_2-s^{(i)}_2 s^{(i)}_1-s^{(2)}_2(s^{(1)}_1-s^{(i)}_1-1)-\\&&-(l_2-s^{(i)}_2+1)(l_1-s^{(i)}_1-s^{(1)}_1+1)\\
  &&\text{as  }l_2-s^{(i)}_2+1\geq 1\\
  0&\leq& t-s^{(d)}_2-s^{(i)}_2 s^{(i)}_1-s^{(2)}_2(s^{(1)}_1-s^{(i)}_1-1)-\\&&-l_2+s^{(i)}_2-1-l_1+s^{(i)}_1+s^{(1)}_1\\
  l_2&\leq& t-s^{(d)}_2-s^{(i)}_2 s^{(i)}_1 -s^{(2)}_2s^{(1)}_1+s^{(2)}_2s^{(i)}_1+s^{(2)}_2+\\&&+s^{(i)}_2-1 +s^{(i)}_1
+s^{(1)}_1-l_1\\
&&\text{we factorize  }-s^{(i)}_2 s^{(i)}_1 +s^{(i)}_1 -s^{(2)}_2s^{(1)}_1+s^{(1)}_1\\
 l_2&\leq& t-s^{(d)}_2-s^{(i)}_1(s^{(i)}_2-1)-s^{(1)}_1(s^{(2)}_2-1)+\\&&
+s^{(2)}_2s^{(i)}_1+s^{(2)}_2+s^{(i)}_2-1-l_1\\
&&\text{we set  }-s^{(i)}_1(s^{(i)}_2-1)+s^{(2)}_2s^{(i)}_1=\\&&=-s^{(i)}_1(s^{(i)}_2-s^{(2)}_2-1)\\
 l_2&\leq& t-s^{(d)}_2-s^{(i)}_1(s^{(i)}_2- s^{(2)}_2-1)-s^{(1)}_1(s^{(2)}_2-1)+\\&&+s^{(2)}_2+s^{(i)}_2-1-l_1\\
 &&\text{we set  }-s^{(1)}_1(s^{(2)}_2-1)+s^{(2)}_2-1=\\&&=-(s^{(1)}_1-1)(s^{(2)}_2-1)\\
 l_2&\leq& t-s^{(d)}_2-s^{(i)}_1(s^{(i)}_2- s^{(2)}_2-1)-\\&&-(s^{(1)}_1-1)(s^{(2)}_2-1)+s^{(i)}_2-l_1\\
 &&\text{if we suppose as first case  }i\neq 2 \text{  then}\\
   l_2&\leq&t-l_1.
\end{eqnarray*}
So we have to consider the case $i=2$ separately.

In this case, having in mind that there exists $s^{(3)}$ (it may be $d=3$) we recalculate
\[|\Delta|\geq s^{(d)}_2+s^{(3)}_2(s^{(2)}_1-1)+s^{(2)}_2(s^{(1)}_1-s^{(2)}_1)\]
and from here
\begin{eqnarray*}
	0&\leq& t-s^{(d)}_2-s^{(3)}_2 (s^{(2)}_1-1)-s^{(2)}_2(s^{(1)}_1-s^{(2)}_1)-\\
	&&-(l_2-s^{(2)}_2+1)(l_1-s^{(2)}_1-s^{(1)}_1+1)\\
	0&\leq& t-s^{(d)}_2-s^{(3)}_2 s^{(2)}_1+s^{(3)}_2-s^{(2)}_2s^{(1)}_1+s^{(2)}_2s^{(2)}_1\\
	&&-l_2+s^{(2)}_2-1-l_1+s^{(2)}_1+s^{(1)}_1\\
    l_2&\leq& t-s^{(d)}_2-(s^{(2)}_1-1)(s^{(3)}_2-s^{(2)}_2-1)-\\&&-(s^{(2)}_2-1)(s^{(1)}_1-2)+2-l_1\\
    l_2&\leq& t-(s^{(d)}_2-2)-(s^{(2)}_1-1)(s^{(3)}_2-(s^{(2)}_2+1))-\\&&-(s^{(2)}_2-1)(s^{(1)}_1-2)-l_1\\
    l_2&\leq&t-l_1,
\end{eqnarray*}
because $s^{(3)}_2-(s^{(2)}_2+1)\geq 0$.

\textbf{Case 1c(ii).} Suppose that $l_1-s^{(i)}_1 < s^{(1)}_1$;  so that the point belongs to  a vertical line that intersects $\Delta$.

Let $j\in \{1,\dots,d\}$ be such that $s^{(j)} \preceq l-s^{(i)}$. Then we add at least $\left(l_2-s^{(i)}_2-s^{(j)}_2+1\right)\left(l_1-s^{(i)}_1-s^{(j)}_1+1\right)$. So,

\begin{eqnarray*}
 0&\leq& t-s^{(d)}_2-s^{(i)}_2 s^{(i)}_1-s^{(2)}_2(s^{(1)}_1-s^{(i)}_1-1)-\\
 &&(l_2-s^{(i)}_2-s^{(j)}_2+1)\left(l_1-s^{(i)}_1-s^{(j)}_1+1\right)\\
 &&\text{since  }\left(l_1-s^{(i)}_1-s^{(j)}_1+1\right)\geq 1\\
 0&\leq& t-s^{(d)}_2-s^{(i)}_2 s^{(i)}_1-s^{(2)}_2(s^{(1)}_1-s^{(i)}_1-1)-l_2+s^{(i)}_2+\\&&+s^{(j)}_2-
\left(l_1-s^{(i)}_1-s^{(j)}_1+1\right)\\
 &&\text{we replace  }-s^{(2)}_2(s^{(1)}_1-s^{(i)}_1-1)\text{  by  }-(s^{(1)}_1-s^{(i)}_1-1)\\
 l_2&\leq& t-s^{(d)}_2-s^{(1)}_1+s^{(i)}_1+1-s^{(i)}_2(s^{(i)}_1-1)+s^{(j)}_2-\\&&-l_1+s^{(i)}_1+s^{(j)}_1-1 \\
 &&\text{we write } 1=-1+2\text{ and factorize  those terms } s^{(i)}_1-1 \\
 l_2&\leq& t-(s^{(i)}_2-2)(s^{(i)}_1-1)+2-s^{(1)}_1-(s^{(d)}_2-s^{(j)}_2)-\\&&-l_1+s^{(j)}_1 \\
 &&\text{we distribute   }2 \text{ and we associate }\\
   l_2&\leq& t-(s^{(i)}_2-2)(s^{(i)}_1-1)-(s^{(d)}_2-s^{(j)}_2-1)-\\&&-(s^{(1)}_1-s^{(j)}_1-1)-l_1
 \end{eqnarray*}

Now, here we have a problem with the bound. If $s^{(i)}_2=1$ then the equality $l_2\leq t-l_1$ does not hold.

To solve this, first note that it must be $i=2$. So, since $f^{(j)}\in F_N$ and we also have $s^{(j)}\preceq l-s^{(i)}$ then the hypothesis of our lemma holds. So that, if $j\neq i$ then we replace $i$ by $j$, and so, we have two options: the first one is that we have $j=1$ which has been already considered, and the other one is that $s^{(j)}_2\geq 2$, and the problem is solved.

It remains to consider the case $i=j=2$ and $s^{(2)}_2=1$. In fact, the problem is that the inequality $l-s^{(k)}\not\in \Delta$ only happens for $k=2$.

So, we may suppose that $s^{(3)}$ exists, because $1<i<d$ and then, we may recalculate a bound for $|\Delta|$; to wit,
\[|\Delta|\geq s^{(d)}_2+(s^{(2)}_1-1)s^{(3)}_2+s^{(2)}_2(s^{(1)}_1-s^{(2)}_1).\]

Now by assummption, $s^{(3)}\not\preceq l-s^{(2)}$; however $s^{(3)}_1\leq l_1-s^{(2)}_1$ and then it must happen that $l_2-s^{(2)}_2<s^{(3)}_2$, so that $l_2-s^{(3)}_2<s^{(2)}_2=1$; hence $l_2 \leq s^{(3)}_2$; that is, the case of Type~\ref{tipos de m}.3.

Frome here, we have

\begin{eqnarray*}
 0&\leq& t-s^{(d)}_2-(s^{(2)}_1-1)s^{(3)}_2-s^{(1)}_1+s^{(2)}_1-\\&&-(l_1-2s^{(2)}_1+1)(l_2-1)\\
 &&\text{as  } (l_1-2s^{(2)}_1+1)(l_2-1)=(l_2-1)+\\
 &&+(l_1-2s^{(2)}_1)(l_2-1)\leq\;l_2-1+l_1-2s^{(2)}_1\\
 0&\leq& t-s^{(d)}_2-(s^{(2)}_1-1)s^{(3)}_2-s^{(1)}_1+s^{(2)}_1-l_2+1-\\&&-l_1+2s^{(2)}_1\\
 0&\leq& t-s^{(d)}_2-(s^{(2)}_1-1)s^{(3)}_2-s^{(1)}_1+s^{(2)}_1-l_2+1-\\&&-l_1+2s^{(2)}_1-2+2\\
 l_2&\leq& t-s^{(d)}_2-(s^{(2)}_1-1)(s^{(3)}_2-2)-s^{(1)}_1+s^{(2)}_1+3-l_1\\
 &&\text{we distribute the summand   }3\\
 l_2&\leq& t-(s^{(2)}_1-1)(s^{(3)}_2-2)-(s^{(d)}_2-2)-\\&&-(s^{(1)}_1-s^{(2)}_1-1)-l_1\\
 &&\text{now we are under the assumption   }s^{(3)}_2\geq 2 \\
 l_2&\leq& t-l_1.
\end{eqnarray*}

\textbf{Case 2.} Suppose that $l_1 < s^{(1)}_1$. 

Note that in this case, it must happen $i\neq 1$ and there is only one posibility; namely, $l_1-s^{(i)}_1 < s^{(1)}_1$; so that the point belongs to a vertical line that intersects $\Delta$.
Then we have that $|\Delta|\geq s^{(d)}_2+s^{(i)}_2 s^{(i)}_1 +s^{(2)}_2(l_1-s^{(i)}_1)$ and we add at least $\left(l_2-s^{(i)}_2-s^{(j)}_2+1\right)\left(l_1-s^{(i)}_1-s^{(j)}_1+1\right)$, where $s^{(j)}\preceq l-s^{(i)}$. We have then

\begin{eqnarray*}
 0&\leq& t-s^{(d)}_2-s^{(i)}_2 s^{(i)}_1-s^{(2)}_2(l_1-s^{(i)}_1)-\\&&-(l_2-s^{(i)}_2-s^{(j)}_2+1)(l_1-s^{(i)}_1-s^{(j)}_1+1)\\
 &&\text{changing  }-s^{(2)}_2(l_1-s^{(i)}_1)\text{  by  }-(l_1-s^{(i)}_1)\\
 &&\text{besides  }\left(l_1-s^{(i)}_1-s^{(j)}_1+1\right)\geq 1\\
 0&\leq& t-s^{(d)}_2-s^{(i)}_2 s^{(i)}_1-l_1+s^{(i)}_1-l_2+s^{(i)}_2+s^{(j)}_2-1\\
 &&\text{grouping and factorizing } \\
 0&\leq& t-s^{(d)}_2-(s^{(i)}_2-1)(s^{(i)}_1-1)-l_1-l_2+s^{(j)}_2\\
 l_2&\leq& t-(s^{(d)}_2-s^{(j)}_2)-(s^{(i)}_2-1)(s^{(i)}_1-1)-l_1\\
 l_2&\leq& t-l_1.
\end{eqnarray*}
\end{proof}

Now, taking the notation of the lemma above, we focus on the situation $f[u]_l\neq 0$, with $l_1+l_2>t$. The easiest cases are $l=(t+a,0)$ and $l=(0,t+a)$, with $a>0$. Here, $f^{(i)}[u]_l\neq 0$ implies that $i=1$ or $i=d$ by Definition~\ref{def recurrencia lineal}. Suppose, first, that $l=(0,l_2)$. By the lemma above, if $f^{(d)}[u]_l\neq 0$ then $l-s^{(d)}\in \Delta$; so that, $l_2< 2s^{(d)}_2\leq 2t$. This means that probably we will need to know all the values $u_l$, with $l=(0,0),\dots,(0,2t-1)$. In an analogous way we may conclude that we will need to know all values $u_l$, with $l=(0,0),\dots,(2t-1,0)$.

The remainder points to be considered are those $l\in \Sigma_{(a,t-a+1)}$, where $0<a\leq t$. Let us show what happen in these case.

\begin{lemma}\label{puntos fuera de la escalera}
 Let $U$ be the syndrome table afforded by $e$ and $\tau$, with $\omega(e)\leq t\leq \lfloor\frac{r_i}{2}\rfloor$, with $i=1,2$. Suppose that, following the BMSa we have constructed, for $u=u^l$ the sets  $\Delta(u)=\Delta$ and $F\in\fF(u)$ and $f[u]_l\neq 0$ with $f\in F$.  If $l=(a,t-a+1)$, where $a\in \Z_{r_1}$ is such that $0< a\leq t$ then one has $l\not\in LP(F)+\Delta$, and hence $\Sigma_{l}\cap (\Delta+LP(F))=\emptyset$.
 
As a consequence, if $n\in \Sigma_{(a,t-a+1)}$, where $a\in \Z_{r_1}$ is such that  $0< a\leq t$ then $f[u]_n= 0$ for any $f\in F$.
\end{lemma}
\begin{proof}
 Set, again, $F=\left\{f^{(1)},\dots,f^{(d)}\right\}$, $LP(f^{(i)})=s^{(i)}$ and $l=(a,t-a+1)$. We shall consider two cases. 
 
\textit{Case 1.} Suppose that $a<s^{(1)}_1$ and $l\in s^{(i)}+\Delta$, for some $i\in \{1,\dots,d\}$. Then, take the maximum of those $k\in \{1,\dots,d\}$ such that $l\preceq s^{(i)}+\left(s^{(k)}_1-1,s^{(k+1)}_2-1\right)$. We have to consider more cases, having in mind the inequality $1\leq a<s^{(i)}_1+s^{(k)}_1$.
 
Suppose that $i=d$. Then, $s^{(k)}_1>1$, and so $C=\left\{\left(s^{(k)}_1-1,0\right),\dots,\left(s^{(k)}_1-1,s^{(k+1)}_2-1\right)\right\}\subset \Delta$, and it cannot be the first vertical line of $\Delta$. Besides, we also have that $V=\left\{(1,0),\dots,(a,0)\right\}\subset \Delta$, since $a<s^{(1)}_1$. From here, we get $t\geq s^{(d)}_2+s^{(k+1)}_2+a-1$, which contradicts that $l\in s^{(i)}+\Delta$.
 
If $i\neq d$ then  $C=\left\{\left(s^{(i)}_1,0\right),\dots,\left(s^{(i)}_1,s^{(i)}_2-1\right)\right\}\subset \Delta$, and it cannot be the first vertical line of $\Delta$.  Besides, we also have that $V=\left\{(1,0),\dots,(a,0)\right\}\subset \Delta$.
 
From here, we have $t\geq s^{(d)}_2+s^{(i)}_2+a-1$, which contradicts that $l\in s^{(i)}+\Delta$.\\
 
\textit{Case 2}. Suppose that $a\geq s^{(1)}_1$. If $a>s^{(1)}_1-1+s^{(j)}_1$ for any $j\in\{1,\dots,d\}$ (equivalently $a>2s^{(1)}_1-1$), then we are done. Othewise, let $i\in \{1,\dots,d-1\}$ such that  $s^{(i+1)}_1+ s^{(1)}_1-1< a \leq s^{(i)}_1+s^{(1)}_1-1$. We are going to see that $t-a\geq s^{(i+1)}_2-1$, which implies that $l\not\in LP(F)+\Delta$, because the rightmost point of $LP(F)+\Delta$ at the same level or higher than $l$ is $(s^{(i+1)}_1+s^{(1)}_1-1,s^{(i+1)}_2)$ and the rest of the points of this set are located under $l$.  Clearly, we are only interested in the case $s^{(i+1)}_2-1>0$.
 
We shall separate the case in which $s^{(1)}_1=1$. Then $i=1$ and $LP(F)+\Delta$ is a rectangle whose point in the upper right corner, the greatest, is $(1,s^{(i+1)}_2-1)=(1,s^{(d)}_2-1)$. If $a>1$ clearly $l\not\in LP(F)+\Delta$. In case $a=1$ we have $l=(1,t)$ and since $s^{(d)}_2-1<t$ we are done.
 
Now, suppose that $s^{(1)}_1>1$. If $s^{(i)}_1=1$ then $i=d-1$ and $a=s^{(1)}_1$; so that $s^{(d)}_2+s^{(1)}_1-1\leq t$, and we are done. Now suppose that $s^{(i)}_1>1$. We shall counting the elements of $\Delta$. The number of points of its first vertical line is $s^{(d)}_2$. We have to add the points of the vertical line $C=\left\{\left(s^{(i)}_1-1,1\right),\dots,\left(s^{(i)}_1-1,s^{(i+1)}_2-1\right)\right\}$, (which is not the first one) and those of the rows $R_0=\left\{\left(1,0\right),\dots,\left(s^{(1)}_1-1,0\right)\right\}$ and, as $s^{(i+1)}_2-1>0$, also $R_1=\left\{\left(1,1\right),\dots,\left(s^{(i)}_1-1,1\right)\right\}$, inside $\Delta$. So that we have counted $t\geq s^{(d)}_2+s^{(i+1)}_2-1+s^{(1)}_1-1+s^{(i)}_1-1$. From here,
 \begin{eqnarray*}
  t-a&\geq&-a+s^{(1)}_1+s^{(i)}_1-1 +s^{(d)}_2-1+s^{(i+1)}_2-1\\
   t-a&\geq&s^{(d)}_2-1+(s^{(1)}_1+s^{(i)}_1-1-a)+s^{(i+1)}_2-1\\
    t-a&\geq& s^{(i+1)}_2-1.
 \end{eqnarray*}
\end{proof}

Let us summarize the results above in the following theorem.

\begin{theorem}\label{puntos fuera de la escalera global}
 Let $U$ be the syndrome table afforded by $e$ and $\tau$, with $\omega(e)\leq t\leq \lfloor\frac{r_i}{2}\rfloor$, with $i=1,2$, as in Definition~\ref{def de U}. Suppose that, following the BMSa we have constructed, for $l=(l_1,l_2)$, and $u=u^l$, the sets  $\Delta(u)=\Delta$ and $F\in\fF(u)$. For any $f\in F$, we have that:
\begin{enumerate}
 \item If $f\in F$ is such that $f[u]_l\neq 0$ and $l\not\in LP(f)+\Delta$ then $l_1+l_2\leq t$.
 \item  If $l_k>2t-1$ then $f[u]_l=0$, for $k\in \{1,2\}$. If $l=(0,l_2)$ or $l=(l_1,0)$ with $l_k\leq 2t-1$ then it may happen that $f[u]_l\neq 0$.
 \item If $l_1,l_2\neq 0$ and $l_1+l_2>t$ then $f[u]_l= 0$ for any $f\in F$, because $l\in \Sigma_{(a,t-a+1)}$ for some $0<a\leq t$.
\end{enumerate}
\end{theorem}

The theorem above suggest us that the following set of indexes is large enough to construct a Groebner basis for $\bL(U)$.

\begin{definition}\label{def conjunto adecuado}
 We shall call $\S(t)$ the subsets of $I$ of the form
 
 \begin{eqnarray*}
 \S(t)=\left\{(0,j)\tq j=0,\dots, 2t-1\right\}\cup\\ \cup \left\{(i,0)\tq i=0,\dots, 2t-1\right\}\cup\\
  \cup \left\{(i,j)\tq i,j\neq 0\;\wedge\; i+j\leq t\right\}.
 \end{eqnarray*}
 In addition,
 \begin{enumerate}
  \item We say that $\S(t)$ satisfies the (lexicographic) l-condition if $u_{(0,j)}\neq 0$, for some $j<t$.
   \item We say that $\S(t)$ satisfies the (graduate) g-condition if $u_{(i,j)}\neq 0$, for some $(i,j)$, with $i+j=1$.
 \end{enumerate}
\end{definition}

In \cite[Corollary, p. 1598]{Blah} Blahut proves that, in the univariate case, if one has an error polynomial satisfying $\omega(e)\leq t$, with known list of syndromes $\{S_0,\dots,S_{2t-1}\}$ (where $S_0\neq 0$) then it is always possible to find the locator polynomial. This suggests that $\S(t)$ is a natural extension of the univariate case.

Let us also comment that similar (but not equal) sets may be found in \cite[p. 1614]{Blah} (see also \cite{saints heegard}) in the context of error correcting codes: an hyperbolic code $C$ of designed distance $d\leq d(C)$ is a bivariate abelian code satisfying that $\A_d=\{(i,j)\tq (i+1)(j+1)\leq d\}\subset \D_\alpha(C)$. As instance, $\A_{10}\varsubsetneq \S(5)$.

The following example shows that we may find problems evaluating some $f[u]_l$. Specifically, it is possible that we may need to use values outside $\S(t)$ in the implementation of the BMSa.

\begin{example}\label{ejemplo del problema de evaluar}
 Set $q=2$, $r_1=5$ and $r_2=7$. So $\L=\F_{2^{12}}$. Let $a\in \L$ be the primitive root of $\L$, with minimal polynomial $m_a=x^{12}+x^{7}+x^6+x^5+x^3+x+1$. Then $\alpha_1=a^{819}$ and $\alpha_2=a^{585}$ are the corresponding primitive roots of unity.
 
 Set $t=2$, $e=X_1^2+X_1X_2$ and $\tau=(0,1)$.

 \[\left(u_n\tq n\in\S(2)\right)=\begin{pmatrix}
        \alpha_2^3 & \alpha_2^6 & \alpha_2 & \alpha_2^5\\
        a^{1513} & a^{3733} \\
        \alpha_1^4 \\
        \alpha_1
      \end{pmatrix}\subset U.\]
Then, we implement the BMSa following the lexicographic order.

\begin{footnotesize}
 \[\begin{array}{|l|l|l|l|}\hline
 l&F\subset \bL(u^{l+1})&G&\Delta(u^{l+1})\\ \hline
  \text{Initializing}&\{1\}&\emptyset&\emptyset\\ \hline
(0,0)\rightarrow&\{X_1,X_2\}&\{1\}&\{(0,0)\}\\ \hline
(0,1)\rightarrow&\{X_1,X_2+\alpha_2^3\}&\{1\}&\{(0,0)\}\\ \hline
 (0,2)\rightarrow&\{X_1,X_2^2+\alpha_2^3X_2+\alpha_2\}&\{X_2+\alpha_2^3\}&\{(0,0),(0,1)\}\\ \hline
 (0,3)\rightarrow &\text{Same}&\text{Same}&\text{Same}\\ \hline
  (1,0)\rightarrow&\parbox{3cm}{$
  \{X_1+a^{3268}X_2^2 +a^{928}X_2,\;X_2^2+\alpha_2^3X_2+\alpha_2\}$}&\{X_2+\alpha_2^3\}&\{(0,0),(0,1)\}\\ \hline
    (1,1)\rightarrow&\parbox{3cm}{$
    \{X_1+a^{3268}X_2^2+a^{1393}X_2+a^{546},\;X_2^2+\alpha_2^3X_2+\alpha_2\}$}&\{X_2+\alpha_2^3\}&\{(0,0),(0,1)\}\\ \hline
\end{array}\]
\end{footnotesize}

Now, at this point, if one wants to evaluate $f^{(1)}[u]_{(2,0)}$, one may see that we need to consider the index $(1,2)$, which is not an element of $\S(t)$. Note that $f^{(1)}$ has a term of multigrade $s^{(2)}=(0,2)$.

Let us replace $F=\{X_1+a^{3268}X_2^2+a^{1393}X_2+a^{546},\;X_2^2+\alpha_2^3X_2+\alpha_2\}$ with $F'=\{X_1+a^{2886}X_2+a^{3349},\;X_2^2+\alpha_2^3X_2+\alpha_2\}$. One may check that it is also a minimal set of polynomials (but now, by Proposition~\ref{forma normal en Lambda} below, in ``normal form'').

Now, to evaluate $(X_1+a^{2886}X_2+a^{3349})[u]_{(3,0)}$ one need the index $(2,1)$, which is not considered, even the elements in $F'$ are in normal form. Note that, setting $l=(3,0)$, $\tau=(0,1)$ and $s=(1,0)$, we have $l-s+m\not\in \S(t)$.
 \end{example}
 
So we identify two different problems during the implementation of the BMSa. Our context is as follows: we have a doubly periodic array, $U$, of period $r_1\times r_2$ and, following the BMSa on a set of indexes $\S(t)$, we get a set $F$ of minimal polynomials (see Definition~\ref{base de u}) for $\bL(u^l)$, with $l\in\S(t)$. Consider $f\in F\in \fF(u^l)$, with $f=\sum_{m\in \supp(f)}f_m\X^m$. Then it may happen that:
 \begin{enumerate}
  \item The polynomial $f$ has a term $f_m$ for which $m\not\in\S(t)$.
  \item Even if $F$ is in normal form, $f$ has a term $f_m$ for which $l-LP(f)+m\not\in \S(t)$.
 \end{enumerate}

The solution of the first problem is given by the following result, which comes directly from Proposition~\ref{combinaciones lineales en lambdas}. It is the key to obtain polynomials equivalent to the normal forms of Groebner basis. 

\begin{proposition}\label{forma normal en Lambda}
 Let $U$ be a doubly periodic array of period $r_1\times r_2$ and $F$ a set of minimal polynomials (see Definition~\ref{base de u}) for $\bL(u^l)$, with $l\in I$. Then, there exists a set $F'$ of minimal polynomials for $\bL(u^l)$ satisfying the following property: for any $f\in F'$ and for all $m\in \supp(f)\setminus\{LP(f)\}$ we have $m \preceq LP(f')$, for all $f'\in F'$; that is, $m\in \Delta(u^l)$.
\end{proposition}

Now we deal with the second problem. The solution will follow from the next technical lemma.

\begin{lemma}\label{delta+delta menor que At}
 Let $\Delta$ be a delta-set with set of defining points $S$. If $|\Delta|\leq t$ then 
 \begin{enumerate}
  \item $\Delta+\Delta \subset \S(t)$.
  \item $S+\Delta \subset \S(t)$.
 \end{enumerate}
\end{lemma}
\begin{proof}
 \textit{1)} Suppose we have two points $(i_1,j_1),(i_2,j_2)\in \Delta$.
 
We separate in cases. The cases where some coordinate of $\left(i_1,j_1\right)$ and some coordinate of $\left(i_2,j_2\right)$ are zero simultaneously are easy to get.
 
 
To get the other cases we only have to consider points of the form $\left(s^{(i)}_1-1,s^{(i+1)}_2-1\right)$ and $\left(s^{(j)}_1-1,s^{(j+1)}_2-1\right)$ with $i\leq j$ and $s^{(k)}_l\geq 1$ for $k\in \{i,i+1,j,j+1\}$ and $l\in \{1,2\}$.
 
First, suppose that $i=j$. Then we may suppose that $s^{(k)}_l\geq 2$ for $k\in \{i,i+1,j,j+1\}$ and $l\in \{1,2\}$, because, otherwise they has been considered. We have that 
\begin{eqnarray} \label{primera desigualdad delta+delta menor que At}\nonumber 
  t&\geq&  s^{(i)}_1s^{(i+1)}_2=\left(s^{(i)}_1-2\right)\left(s^{(i+1)}_2-2\right)+\\ \nonumber
  &&+2s^{(i)}_1+2s^{(i+1)}_2-4\\ &\geq& 2s^{(i)}_1+2s^{(i+1)}_2-4.
\end{eqnarray}

Now, suppose $i<j$. In this case 
 \begin{eqnarray}\label{segunda desigualdad delta+delta menor que At}\nonumber
  t&\geq& s^{(j)}_1s^{(j+1)}_2+s^{(i+1)}_2\left(s^{(i)}_1-s^{(j)}_1\right)\\ \nonumber
  &\geq& s^{(j+1)}_2+\left(s^{(j)}_1-1\right)s^{(j+1)}_2+s^{(i+1)}_2s^{(i)}_1-s^{(i+1)}_2s^{(j)}_1\\ \nonumber
  &\geq& s^{(j+1)}_2+\left(s^{(j)}_1-1\right)\left(s^{(j+1)}_2-s^{(i+1)}_2\right)+\\\nonumber &&+s^{(i+1)}_2s^{(i)}_1-s^{(i+1)}_2\\  \nonumber
  &\geq& s^{(j+1)}_2+\left(s^{(j)}_1-1\right)\left(s^{(j+1)}_2-s^{(i+1)}_2\right)+\\\nonumber &&+\left(s^{(i+1)}_2-1\right)\left(s^{(i)}_1-1\right)+s^{(i)}_1-1\\  \nonumber
  &\geq& s^{(j+1)}_2+\left(s^{(j)}_1-1\right)+\left(s^{(i+1)}_2-1\right)+s^{(i)}_1-1\\  
   &\geq & s^{(j+1)}_2+s^{(j)}_1+s^{(i+1)}_2+s^{(i)}_1-3.   
 \end{eqnarray}

  \textit{2)} We consider the points of the form $\left(s^{(i)}_1-1,s^{(i+1)}_2-1\right)$ and $\left(s^{(j)}_1,s^{(j)}_2\right)$ with $i\leq j$. The other cases are analogous.
  
  If $i=j$, in view of Inequality \eqref{primera desigualdad delta+delta menor que At} we only have to analize the case $t = s^{(i)}_1s^{(i+1)}_2$, but it happens only if $i=1$ and $d=i+1=2$. For $\left(s^{(1)}_1-1,s^{(2)}_2-1\right)$ and $\left(0,s^{(2)}_2\right)$ we have, for $s^{(2)}_2\geq 2$, 
\[ 2s^{(2)}_2+s^{(1)}_1-2\leq 2s^{(2)}_2+2s^{(1)}_1-4\leq t, \;\text{ by Inequality } \eqref{primera desigualdad delta+delta menor que At}\]
and the case $s^{(2)}_2=1$ get $s^{(1)}_1\leq t$.

If $i<j$, having in mind that $s^{(j)}_2\leq s^{(j+1)}_2-1$ we have
\begin{eqnarray*}
 s^{(j)}_2+s^{(j)}_1+s^{(i+1)}_2+s^{(i)}_1-2\leq (s^{(j+1)}_2-1)+s^{(j)}_1+\\+s^{(i+1)}_2+s^{(i)}_1-2 \leq t,\;\text{ by Inequality } \eqref{segunda desigualdad delta+delta menor que At}
\end{eqnarray*}
\end{proof}

Now we describe how to proceed to solve the second problem once the 
elements of $F$ are written in normal form (by Proposition~\ref{forma normal en Lambda}). Suppose that for $f\in F$, with $s=LP(F)$, in the computation of $f[u]_l$ it happens that $l-s+m \not\in \S(t)$, for some $m\in\supp(f)$ with $m\in \Delta$ or $m=s$. Then, by Lemma~\ref{delta+delta menor que At} we have that $l-s\not\in \Delta$; moreover, for any $\Delta'\supseteq \Delta$, with $|\Delta'|\leq t$, we also have that  $l-s\not\in \Delta'$; in particular $l-s\not\in \Delta(u^{l+1})$, hence, it must happen that $f[u]_l=0$.\\

Let $U$ be the syndrome table afforded by $e$ and $\tau$, with $\omega(e)\leq t$, as in Definition~\ref{def de U}. The following result guarantees that the set of polynomials obtained at the last iteration in the BMSa over a set of the form $\S(t)$ is exactly a Groebner basis for $\bL(U)$,

\begin{theorem}\label{condicion suficiente}
Let $U$ be the syndrome table afforded by $e$ and $\tau$, with $\omega(e)\leq t\leq \lfloor\frac{r_i}{2}\rfloor$, for $i=1,2$. It is possible to find a Groebner basis for the ideal $\bL(U)$ following the BMSa by using only the values $\{u_n\tq n\in \S(t)\}$ with either the lexicographic ordering, with the l-condition, or the graduate ordering, with the g-condition.

Hence, it is possible to find $\supp(e)$, from a defining set of $L(e)$.
\end{theorem}
\begin{proof}
We begin by considering the lexicographic ordering, $X_1>X_2$. First, note that while the implementation of the algorithm, we always have that $f^{(1)}\in F_N$ implies $l_1\geq s^{(1)}_1$ and the form of the new defining points in Theorem~\ref{tipos de m}, drives us to the inequality $l_1\geq s^{(1)}_1-1$. 
 
At initializing the algorithm one takes $F=\{1\}$, so  that $1[u]_{(0,j)}=u_{(0,j)}$ and the first two defining points obtained are always $(1,0)$ and $(0,j+1)$, which means that it is essential that  $u_{(0,j)}\neq 0$ for some $j<t$.  After that, there are no restrictions.
 
Making the algorithm on the first vertical line ($l=(0,b)$, with $b\geq 0$) we know that we have to compute at most the points $l=(0,j)$, with $j=0,\dots,2t-1$ and we also know that the polynomials obtained are valid for all values in the steps corresponding to the first vertical line. Now, suppose that we have calculated $\Delta$ for the points $l=(l_1,l_2)$, with $l_1>0$, $l_1\geq s^{(1)}_1-1$ and $l_2=0,\dots,t-l_1$. Then for any incoming point $l$ over such vertical line we have that $f^{(i)}[u]_l=0$, because, otherwise,  Theorem~\ref{puntos fuera de la escalera global} says us that hypothesis of Lemma~\ref{cota para l2} are satisfied; but $l_2>t-l_1$, which is a contradiction. So we can ensure that the constructed polynomials are valid up to $(l_1,r_2)$; that is, in all the vertical line.

The last point in which  $\Delta$ may be increased is $l=(t,0)$. After it, Lemma~\ref{cota para l2} and Theorem~\ref{puntos fuera de la escalera global} assure us that there are not increasing of the area; however, any point of the form $l=(l_1,0)$, with $l_1\leq 2t-1$, may verify that $l\in LP(F)+\Delta$, and then the set $F$ may be modified by Procedure~\ref{proc 1}. For this reason it is necessary to consider all of them.
 
For the points $l\geq_T (2t,0)$ it happens that $l\not\in LP(F)+\Delta$ and, besides this,  $l_1+l_2>t$. Hence it must happen that for all $f\in F$, the equality $f[u]_l=0$ holds.
 
Now let us consider the graduate ordering, together with $X_1>X_2$. Suppose we have constructed the minimal sets through the values $l$, following the diagonals in $\Sigma_0$, until we reach the diagonal $t$; that is $\{l=(l_1,l_2)\tq l_1+l_2=t\}$ and  call $F$ the last minimal set, with $\Delta= \Delta(F)$. Consider a point $l=(l_1,l_2)$ such that $l_1+l_2\geq t+1$. If  one has that $l_1,l_2\neq 0$ then Theorem~\ref{puntos fuera de la escalera global} says that  $f[u]_l=0$ for all $f\in F$.
Finally, it may happen that $f[u]_l\neq 0$ for $l=(a,0),(0,a)$ with $a \in \{t+1,\dots,2t-1\}$ (the cases $l=(j,0)$, with $j\geq 2t$ has been already seen). We will continue forming minimal sets of polynomials until consider all of them.
 
Thus, in any of the two monomial orderings, the polynomials of $F$ are valid over all $\Z_{r_1}\times \Z_{r_2}$ and so $F\subset \bL(U)$. Hence $\langle F \rangle=\bL(U)$. By applying Theorem~\ref{igualdad ideales} we get the result. 
\end{proof}

\begin{remark}\label{Ejemplo extremo Lex}
 The inclusion in $\S(t)$ of points of the form $(a,0)$ and $(0,a)$ for $t+1\leq a\leq 2t-1$ is not superfluous, as we may see in the following example that we comment briefly.
 
 Set $q=2$ and $r_1=r_2=15$. Then $\L=\F_{2^4}$. Let $e=X_1^8+X_1^4+X_1^2+X_1$; so that $t=4$ and set $\tau=(3,0)$; that is $u_{(0,0)}=e(\overline{\alpha}^{(3,0)})$. The reader may check that in the step $l=(6,0)$ one get $F=\{X_1^4+X_1^3+X_1^2,X_2+1\}\subset \bL(u^{(7,0)})$, and, the last step, $l=(7,0)$ produces the Groebner basis $F=\{X_1^4+X_1+1,X_2+1\}$ for $\bL(U)$.\\
 
 The same happens in the univariate case. For $e=x^8+x^4+x^2+x$ and a fixed root of unity $\alpha\in\L$, one may check that, by applying the BMa \cite[p. 1596]{Blah} one needs to consider all syndromes: $S_0=e(\alpha^3),\dots,S_7=e(\alpha^{10})$, to get the locator polynomial $\Lambda^{(8)}(x)=1+x+x^4$.
\end{remark}

 \begin{example}\label{Ejemplo 5 por 5 xy^2+x^2y^2}
 Set $q=2$, $r_1=r_2=5$, $a\in \F_{2^4}$ be the primitive root associated to $x^4+x+1$, and $\bar\alpha=(a^3,a^3)$. Let $e=X_1X_2^2+X_1^2X_2^2$ (so $t=2$) and $\tau=(1,1)$. For example, the first value $u_{(0,0)}=e(\bar\alpha^{(1,1)})=a^{8}$ and the last one $u_{(3,0)}=e(\bar\alpha^{(4,1)})=a^{14}$. So that we arrange
  \[\left(u_n\tq n\in\S(2)\right)=\begin{pmatrix}
        a^{8} & a^{14} & a^{5}& a^{11}\\
        a^{10} & a \\
        a^{7} \\
        a^{14}
      \end{pmatrix}\subset U.\]

Next table summarizes all computation with respect to the lexicographic ordering.
\begin{footnotesize}
 \[\begin{array}{|l|l|l|l|}\hline
 l&F\subset \bL(u^{l+1})&G&\Delta(u^{l+1})\\ \hline
  \text{Initializing}&\{1\}&\emptyset&\emptyset\\ \hline
(0,0)\rightarrow&\{X_1,X_2\}&\{1\}&\{(0,0)\}\\ \hline
(0,1)\rightarrow&\{X_1,X_2+a^{6}\}&\{1\}&\{(0,0)\}\\ \hline
 (0,2),(0,3)\rightarrow&\text{Same}&\text{Same}&\text{Same}\\ \hline
(1,0)\rightarrow& \{X_1+a^2,X_2+a^{6}\} &\{1\}&\{(0,0)\}\\ \hline
(1,1)\rightarrow &\text{Same}&\text{Same}&\text{Same}\\ \hline
(2,0)\rightarrow& \parbox{2.3cm}{$
\{X_1^2+a^2X_1+a^9,X_2+a^{6}\}$} &\{X_1+a^2\}&\{(0,0),(0,1)\}\\ \hline
(3,0)\rightarrow& \text{Same} & \text{Same} & \text{Same} \\\hline
\end{array}\]
\end{footnotesize}

Now one may check that \newline $\D_{\bar\alpha}\left(\{X_1^2+a^2X_1+a^9,X_2+a^{6}\}\right) =\{(1,2),(2,2)\}=\supp(e)$.
\end{example}

The reader may see in the following remark that even if $l$-condition is not satisfied, the procedure may works. An interesting problem is to improve both conditions.

\begin{remark}\label{puede haber muchos ceros}
 In the setting of the example above, one may check that, if $\tau=(1,1)$ is replaced by $\tau=(0,1)$ then one obtains
  \[\left(u_n\tq n\in\S(2)\right)=\begin{pmatrix}
  0 & 0 & 0 & 0\\
        a^{8} & a^{14}\\
        a^{10} \\
        a^{7}
      \end{pmatrix}\subset U\]
and one may check that the same Groebner basis may be obtained by applying the BMSa.
\end{remark}

Once we have obtained a Groebner basis for $\bL(U)$, we may know any of the values $u_n\in U$ and all coefficients of $e\in \F(r_1,r_2)$.

\begin{corollary}\label{completando U y e}
 In the setting of the theorem above it is possible to find:
 \begin{enumerate}
  \item Any value $u_n\in U$, with $n\in \Sigma_0$.
  \item Any value $e_s\in \F$, with $s\in \supp(e)$.
 \end{enumerate}
\end{corollary}
\begin{proof}
 \textit{1)} We may compute all values $u_n$ with $n\in \Z_{r_1}\times \Z_{r_2}$ as in \cite[Lemma 10]{Sakata 2}; that is, for those unknown values $u_n$ with $n\not\in \Delta(U)$, we consider $X^n$ and we write $X^n=h^{(n)}+R_n$, where $h^{(n)}$ is the unique linear combination of the minimal Groebner basis for $\bL(U)$, obtained from theorem above. So that, $X^n-R_n=h^{(n)} \in \bL(U)$ and $\supp(R_n)\subset \Delta$, and so we may compute explicitly the values $u_n$ because $h^{(n)}[U]_n=0$. 
 
 \textit{2)} It is clear that, once we have enough values of $U$, we may form a uniquely determined system of equations and that the coefficients of $e$ form a solution.
\end{proof}

We want to rephrase the theorem above by replacing the existence of the polynomial $e\in \L(r_1,r_2)$ by other condition. By the definition of doubly periodic arrays and Remark~\ref{hechos sobre lambda de U}, the ideal $\bL(U)$ always exists; however, there are doubly periodic arrays $U$, of period $r_1\times r_2$, for which $|\Delta(U)|> \lfloor\frac{r_i}{2}\rfloor$ for $i=1,2$ (for example, take a syndrome table $(u_n)$ as in the theorem above, change any of its values $u_n$, with $n\not \in \S(t)$ and apply Corollary~\ref{unicidad arreglo U} below). This is the key to establish the new hypothesis in the following result. 

\begin{theorem}\label{condicion suficiente no locator}
Let $t,r_1,r_2\in \N$, with $t\leq\lfloor\frac{r_i}{2}\rfloor $ for $i=1,2$, and let $U$ be a doubly periodic array of period $r_1\times r_2$, constructed by linear recurring relations and some initial values, such that, $|\Delta(U)|\leq t$. Then, it is possible to find a Groebner basis for $\bL(U)$ following the BMSa by using only the values $\{u_n\tq n\in \tau+\S(t)\}$, for some $\tau\in I$, and $\S(t)$  with either the lexicographic ordering, with the l-condition, or the graduate ordering, with the g-condition.

Moreover, it is possible to find any value $u_n\in U$, with $n\in \Sigma_0$, by using the obtained linear recurring relations.
\end{theorem}
\begin{proof}
 Copy the proofs of the theorem and the first part of the corollay above and paste them here.
\end{proof}

It is easy to see that any doubly periodic array $U$ afforded by  $e \in \L(r_1,r_2)$ with $\omega(e)\leq t\leq\lfloor\frac{r_i}{2}\rfloor $, for $i=1,2$, may be afforded by another polynomial having the same weight and another paremeter $\tau\in I$; for example, the syndrome table in Example~\ref{Ejemplo 5 por 5 xy^2+x^2y^2} above may be also seen as afforded by $e'=a^9X_1X_2^2+a^{12}X_1^2X_2^2$ and $\tau=(0,0)$. The following result is a direct consequence of the previous ones and it tells us that, for a fixed $\tau\in I$ and $t\leq\lfloor\frac{r_i}{2}\rfloor$ the affording polynomial $e$ is unique.

\begin{corollary}\label{unicidad polinomilo e}
 Let $t,r_1,r_2\in \N$, with $t\leq\lfloor\frac{r_i}{2}\rfloor $ for $i=1,2$. Let $e\in \F(r_1,r_2)$ and $\tau\in I$, with $\omega(e)\leq t$. If there exists  $e'\in \F(r_1,r_2)$ with $\omega(e')\leq t$ and $e\left(\overline{\alpha}^{n+\tau}\right)= e'\left(\overline{\alpha}^{n+\tau}\right)$ for all $n\in\S(t)$ then $e=e'$.
\end{corollary}

By replacing the existence of the polynomial by the condition $|\Delta(U)|\leq t\leq\lfloor\frac{r_i}{2}\rfloor $ we may also prove a uniqueness result for doubly periodic arrays.

\begin{corollary}\label{unicidad arreglo U}
 Let $t,r_1,r_2\in \N$, with $t\leq\lfloor\frac{r_i}{2}\rfloor$, for $i=1,2$. Let $U=(u_n)$ be a doubly periodic array of period $r_1\times r_2$, with $|\Delta(U)|\leq t$. If there exists another a doubly periodic array, $V=(v_n)$, of period $r_1\times r_2$, with $|\Delta(V)|\leq t $  such that $u_n=v_n$ for all $n\in\S(t)$ then $U=V$.
\end{corollary}

Our results above may be applied to determine if a  data array $S$ of size $r_1\times r_2$ for which we only know some values is, in fact, a syndrome table generated by a polynomial $e\in \F(r_1,r_2)$, with $\omega(e)=t\leq\lfloor\frac{r_i}{2}\rfloor$.

\begin{corollary}\label{tabla generica para ABMS}
Let $t,r_1,r_2\in \N$, with $t\leq\lfloor\frac{r_i}{2}\rfloor $ for $i=1,2$. Suppose $S$ is a data array of size $r_1\times r_2$ for which we only know some values. Let $T=\{n\in I\tq s_n \text{ is a known value}\}$ and suppose $\tau+\S(t)\subset T$ for $\S(t)$ with either the l-condition, or the g-condition, and some $\tau\in I$.

If there exists a polynomial $e \in \F(r_1,r_2)$ with $\omega(e)\leq t$ such that $e\left(\overline\alpha^n\right)=s_n$ for all $n\in T$ then we may find it by following the BMSa.
\end{corollary}
\begin{proof}
If such a polynomial exists then there exists, in turn, the syndrome table afforded by $e$ and $\tau$, say $U$. Now, for any $m\in \S(T)$ such table verifies that $u_m=s_{\tau + m}$, with $\tau + m\in T$; so that, $u_m$ is a known value. We apply the BMSa and by Corollary~\ref{completando U y e}.\textit{2)} and Corollary~\ref{unicidad arreglo U} we are done.
\end{proof}

Another interesting application is the following termination criteria.

\begin{corollary}
Let $U$ be the syndrome table afforded by $e\in\F(r_1,r_2)$ and $\tau$. Suppose that, following the BMSa we have constructed, for $u=u^l$ the sets  $\Delta(u)=\Delta$ and $F\in\fF(u)$. We also suppose that we have constructed (solving, if it is possible, the corresponding system of equations) a polynomial $e_F \in\F(r_1,r_2)$ such that $\supp(e_F)=\D_{\bar{\alpha}}(F)$. If $e_F\left(\overline{\alpha}^{n+\tau}\right)= u_n$ for all $n\in\S(t)$ then $e=e_F$. Moreover, the normal form of $F$ is a Groebner basis for $\bL(U)$.
\end{corollary}

Note that, in the binary case; that is, if we assume that $e\in \F_2(r_1,r_2)$ it is easy to implement because we have not to solve any system of equations.\\

We finish this section with a comment on complexity of the BMSa. For any $l\in I$ we denote $|l|=|\Sigma_0^{l+1}|$. In \cite[Theorem 3, p. 333]{Sakata 2} it is proved that for any $l\in I$ the complexity of the algorithm in the step $l$ is $\mathcal{O}(|l|^2)$. To complete this result we may compute the maximum number of global steps (see \cite[Exercise 10.9]{Cox et al Using}). By Theorem~\ref{condicion suficiente}, it is $|\mathcal{S}(t)|=\frac{t^2+7t}{2}-1$.

\section{Applications to locator decoding in abelian codes.}

Let us comment briefly the framework of locator decoding in cyclic and abelian codes. A nice and extensive exposition of this topic may be found in \cite{Blah} (see also \cite{Sakata 6}). In the case of cyclic codes, the use of the Berlekamp-Massey algorithm in the context of locator decoding is based on the BCH bound, which may be described in terms of the defining set of a cyclic code. Many algorithms from locator decoding depend on the properties of the discrete Fourier transform, through which syndrome vectors are constructed. We access the values of this vector when the defining set contains a set of consecutive zeros. In the case of abelian codes, the BMSa is also based on the use of the Fourier transform in the construction of syndrome tables (instead of syndrome vectors). As it is done for cyclic codes, the access to the values of the table is made from the defining set; however, in this case we have to replace the use of consecutive zeros and the BCH bound by others tools, and they should work as natural extensions. 

As we commented in Introduction, all previous results on the implementation of the BMSa, from general abelian codes to AG-codes (see \cite[Chapter 10]{Cox et al Using},\cite[Theorem 4.9]{Hackl}, \cite{rubio}, \cite[Section 6]{Sakata 2}, \cite{Sakata 5}, \cite{Sakata 6}), are based on the computation of enough steps to ensure that the size of the set determined by the footprint of $\mathbf{\Lambda}(U)$ or $\Delta(e)$ cannot be increased. Those steps had to be executed and analized one by one.   The goal of our paper is to show that one has to update the minimal set of polynomials only at those indexes contained in the set $\mathcal{S}(t)$ and moreover, none of them may be ignored \textit{a priori}, as it is shown in Remark~\ref{Ejemplo extremo Lex}.

The lower bound for the minimum distance of abelian codes that we shall consider is the multivariate BCH bound, based on the (strong) apparent distance (see \cite{Camion,BBCS2,BGS}). It is easy to prove that any abelian code, $C$, for which $\tau+\S(t)\subset \D_{\overline{\alpha}}(C)$ verifies that its strong apparent distance satisfies $sd^*(C)\geq 2t+1$.

Throughout this section we fix an element $\overline{\alpha}\in \R$. Now we shall apply the results above to locator decoding in abelian codes. A \textbf{bivariate code}, or 2-dimensional abelian code, of length $r$ (see \cite{Imai}) is an ideal in the algebra   $\F(r_1, r_2)$, where $r=r_1r_2$; so we keep all notation and basic results from the Preliminaries.

Let $C$ be an abelian code  in $\F(r_1, r_2)$ with defining set $\D_{\bar\alpha}(C)$ for  $\bar\alpha \in \R$. We recall that for any $f\in\F(r_1, r_2)$ we have $f\in C$ if and only if $f\left(\bar\alpha^n\right)=0$ for all $n\in \D_{\bar\alpha}(C)$. Set $d=d(C)$; the minimum distance and $\delta\leq d$ a lower bound for it; as, for example, the (strong) apparent distance for abelian codes (see \cite{Camion}, \cite{BBCS2} or \cite{BGS}) and $t=\lfloor\frac{\delta-1}{2}\rfloor$ (the bound for) the error capability of $C$.

We also recall the extension of the concept of $q$-cyclotomic coset of an integer to two components. 

\begin{definition}\label{qorbita}
Given an element $(a_1,a_2)\in I$, we define its \textit{$q$-orbit} modulo  $\left(r_1,r_2\right)$ as
	\begin{eqnarray*}
	  Q(a_1,a_2)&=&\left\{\left(a_1\cdot q^i ,a_2\cdot q^i  \right)\tq i\in \N\right\} \subseteq I= \Z_{r_1}\times\Z_{r_2}.
	\end{eqnarray*}
\end{definition}
	
It is easy to see that for every abelian code $C\subseteq\F(r_1,r_2)$, $\D_{\overline{\alpha}}\left(C\right)$ is closed under multiplication by $q$ in $I$, and then $\D_{\overline{\alpha}}(C)$ is necessarily a disjoint union of $q$-orbits modulo $(r_1,r_2)$. Conversely, every union of $q$-orbits modulo $(r_1,r_2)$ defines an abelian code in $\F(r_1,r_2)$. For the sake of simplicity we only write $q$-orbit, and the tuple of integers will be clear from the context.

Suppose a codeword $c\in C$ was sent and we have received a polynomial $f(X_1,X_2)=c(X_1,X_2)+e(X_1,X_2)$; where $\omega(e)\leq t$. We want to determine the error polynomial by using the locator decoding technique (see \cite{Blah,Hackl,Sakata}); that is, we want to construct the error locator ideal $L(e)$, and then, recovering enough syndrome values, determine the coefficients of $e$ in the non-binary case.

\begin{theorem}\label{Teo alg BMS en codigos}
 Let $C$ be an abelian code in $\F(r_1,r_2)$ with correcting capability $t\leq \lfloor\frac{r_i}{2}\rfloor$ (the true capability or a bound), for $i=1,2$. Suppose that there are a fixed $\bar\alpha\in\R$ and $\tau\in I$ such that $\tau+\S(t)\subset \D_{\bar\alpha}(C)$ with $\S(t)$ satisfying the l-condition or the g-condition. Then, in any transmision with no more than $t$ errors, we may decode succesfully by the BMSa.
\end{theorem}
\begin{proof}
 Suppose a codeword $c\in C$ was sent and we have received a polynomial $f(X_1,X_2)=c(X_1,X_2)+e(X_1,X_2)$; where $\omega(e)\leq t$. By definition, for any $n\in \D_{\bar\alpha}(C)$ we have that $f(\bar\alpha^n)=c(\bar\alpha^n)+e(\bar\alpha^n)=e\left(\bar\alpha^n\right)$; so, for all $n\in \D_{\bar\alpha}(C)$, we know the values of the syndrome table $S=\left(s_n\right)_{n\in \Sigma_0}$ of $e(\bar\alpha^n)=s_n$. At this point, we arrive to the framework of Section~\ref{el ideal locator} and Section~\ref{seccion de condiciones suficuentes}. If there exists an element $\tau\in I$ and a set of the form $\S(t)$ as in Definition~\ref{def conjunto adecuado} such that $\tau+\S(t)\subset \D_{\bar\alpha}(C)$, we may define, as in Definition~\ref{def de U} a syndrome table afforded by $e$ and $\tau$, and then, by  Theorem~\ref{condicion suficiente}  we may find $L(e)$, and by Corollary~\ref{completando U y e} we may find the error polynomial $e$.
\end{proof}

\begin{example}\label{ejemplo de (0,t+j)}
 Consider the code $C$, in $\F_2(5,15)$ with primitive root $a$, and $\D_{(\alpha,\beta)}(C)=Q(0,1)\cup Q(1,1)\cup Q(2,1)\cup Q(3,1)\cup Q(4,1)\cup Q(2,3)$. One may check that the strong apparent distance  $sd^*(C)=6$, so that $t=2$ is a lower bound for the error correction capability of $C$. For the error polynomial $e=X_2^2+X_1X_2^3$ and $\tau=(2,1)$ we have the first value $u_{(0,0)}=e(\alpha^2,\beta)=a^{11}$ and the last one $u_{(3,0)}=e(\alpha^0,\beta^1)=a^6$. So that we arrange
  \[\left(u_n\tq n\in\S(2)\right)=\begin{pmatrix}
        a^{11} & a^{6} & a^{13}& a^{13}\\
        a^{7} & a \\
        a^{8} \\
        a^6
      \end{pmatrix}.\]

Next table summarizes all computation with respect to the lexicographic ordering.
\begin{footnotesize}
 \[\begin{array}{|l|l|l|l|}\hline
 l&F\subset \bL(u^{l+1})&G&\Delta(u^{l+1})\\ \hline
  \text{Initializing}&\{1\}&\emptyset&\emptyset\\ \hline
(0,0)\rightarrow&\{X_1,X_2\}&\{1\}&\{(0,0)\}\\ \hline
(0,1)\rightarrow&\{X_1,X_2+a^{10}\}&\{1\}&\{(0,0)\}\\ \hline
 (0,2)\rightarrow& \parbox{2.3cm}{$
 \{X_1,X_2^2+a^{10}X_2+a\}$}&\{X_2+a^{10}\}&\{(0,0),(0,1)\}\\ \hline
  (0,3)\rightarrow& \parbox{2.3cm}{$
  \{X_1,X_2^2+a^{6}X_2+a^5\}$}&\{X_2+a^{10}\}&\{(0,0),(0,1)\}\\ \hline
(1,0)\rightarrow& \parbox{2.3cm}{$
                  \{X_1+a^2X_2+1,X_2^2+a^{6}X_2+a^5\}$}
&\{X_2+a^{10}\}&\{(0,0),(0,1)\}\\ \hline
(1,1)\rightarrow &  \parbox{2.3cm}{$
\{X_1+a^8X_2+a^5,\\
X_2^2+a^{6}X_2+a^5\}$}
&\{X_2+a^{10}\}&\{(0,0),(0,1)\}\\ \hline
  (2,0),(3,0)\rightarrow&
\text{Same} & \text{Same} & \text{Same}
\\\hline
\end{array}\]
\end{footnotesize}
The reader may check that $\D_{\bar\alpha}(\bL(U))=\D_{\bar\alpha}(\langle F\rangle)=\{(0,2),\;(1,3)\}$.
\end{example}

One may check that there exist abelian codes having strong apparent distance $sd^*(C)\geq 2t+1$ but they have not any set of the form $\tau+\S(t)$ in their defining sets. We finish by relating our results with the  bivariate BCH bound (see \cite{BBCS2}) in abelian codes.\\

Recall from \cite[Theorem 30, Corollary 31]{BBCS2} the BCH multivariate bound theorem in the case of two variables. Let $C$ be a nonzero bivariate abelian code in $\F(r_1,r_2)$. Define the sets $\tilde\gamma \subseteq \{1,2\}$, $\tilde\delta=\{\delta_k\tq k\in \tilde\gamma \text{ and } 2\leq  \delta_k\leq r_k\}$ and a list of integers $\hat b=(b_k)_{k\in\bar\gamma}$ with $b_k\geq 0$. For each $k\in \tilde\gamma$ consider the list of consecutive integers modulo $r_k$, $J_k=\left\{\overline{b_k},\dots,\overline{b_k+\delta_k-2}\right\}$, where $\overline{b_k+c}$ means the canonical representative of $b_k+c\in \Z$, modulo $r_k$, and set $A_k=\left\{(i_1,i_2)\in I\tq i_k\in J_k\right\}$. If $\cup_{k\in\tilde\gamma}A_k\subset \D_{\overline{\alpha}}(C)$ for some $\overline{\alpha}\in \R$, then the minimum distance $d(C)\geq \prod_{k\in\tilde\gamma}\delta_k$.

Now, the following lemma is immediate.

\begin{lemma}\label{t y la cota BCH}
 Let $C$ be a nonzero abelian code. Let $\tilde\gamma \subseteq \{1,2\}$, $\tilde\delta=\{\delta_k\tq k\in \tilde\gamma \text{ and } 2\leq  \delta_k\leq r_k\}$ and $\hat b=(b_k)_{k\in\bar\gamma}$ with $b_k\geq 0$. For each $k\in \tilde\gamma$ consider the list of consecutive integers modulo $r_k$, $J_k=\left\{\overline{b_k},\dots,\overline{b_k+\delta_k-2}\right\}$ and set $A_k=\left\{(i_1,i_2)\in I\tq i_k\in J_k\right\}$, as above. If $\cup_{k\in\tilde\gamma}A_k\subset \D_{\overline{\alpha}}(C)$ for some $\overline{\alpha}\in \R$, then $\D_{\overline{\alpha}}(C)$ contains a set of the form $\tau+\S(t)$ for $t$ satisfying:
 \begin{enumerate}
  \item  $t\leq \min\{\lfloor \frac{\delta_k-1}{2}\rfloor, \lfloor \frac{r_i}{2}\rfloor,\;i=1,2\}$, in case $\tilde\gamma=\{k\}$.
  \item $t\leq\min\{\delta_1+\delta_2-3, \lfloor \frac{r_i}{2}\rfloor,\;i=1,2\}$, in case $\tilde\gamma=\{1,2\}$.
 \end{enumerate}

\end{lemma}
\begin{proof}
 By definition of the sets $A_k$, for $k=1,2$ it is clear that one may take $m=(b_1,0)$ if $\tilde\gamma=\{1\}$, $m=(0,b_2)$ if $\tilde\gamma=\{2\}$ and $m=(b_1,b_2)$ if $\tilde\gamma=\{1,2\}$.
\end{proof}

Note that, the restriction $t\leq \lfloor \frac{r_i}{2}\rfloor$, for $i=1,2$, must be imposed in order to implement the BMSa.

\begin{proposition}
 Notation as in Lemma~\ref{t y la cota BCH}.  Let $C$ be a nonzero abelian code satisfying all conditions in such lemma for the integer $t$, with $\S(t)$ satisfying the $l$-condition or the $g$-condition. Then in any transmision with no more than $t$ errors, we may decode succesfully by the BMSa.
\end{proposition}

Let $\tilde\gamma$, $\tilde\delta$ and $\hat b$, as above. For $k\in \tilde{\gamma}$, we define $I(k,u)=\{\left(i_1,i_2\right)\in I\tq i_k=u\}$. Recall from \cite[Definition 42]{BGS} that an abelian code $C$ in $\F(r_1,r_2)$ is a bivariate BCH code of designed distance $\tilde\delta$ if for some $\overline{\alpha} \in U$,
\[\D_{\overline{\alpha}}(C)=\bigcup_{k\in\bar\gamma} \bigcup_{l=0}^{\delta_k-2}\bigcup_{(i_1,i_2)\in I(k,\overline{b_k+l})}Q((i_1,i_2)).\]

The BCH bivariate (and multivariate) codes are denoted by $B_q(\overline{\alpha},\tilde\gamma,\tilde\delta,\hat b)$. In \cite{BGS} true minimum distance for abelian codes was studied; in particular for BCH bivariate codes (see \cite[Theorem 44]{BGS}). We may apply the proposition above to see that for $t$ satisfying the conditions in Lemma~\ref{t y la cota BCH}, then in any transmision with no more than $t$ errors, we may decode succesfully by the BMSa.

\section{Conclusions}
In this paper, we have studied the problem of the computation of Groebner basis for the ideal of linear recurring relations of a doubly periodic array. We have present a set of indexes such that, guarantees that the set of polynomials obtained at the last iteration in the Berlekamp-Massey-Sakata algorithm is exactly a Groebner basis for the mentioned ideal. We also  applied this to improve locator decoding in abelian codes.


\ifCLASSOPTIONcaptionsoff
  \newpage
\fi

\begin{IEEEbiographynophoto}{J. J. Bernal}
 was born in Murcia, Spain, in May 1976. He received the B.S. degree in Mathematics in 1999 and the M.S. degree in Advanced Mathematics in 2007 from the University of Murcia. He received the Ph. D. degree in 2011 from the University of Murcia.
 
 He has been an Associate Professor in the Applied Mathematics Department
of the Faculty of Computer Sciences, in the Quantitative Methods
Department of the Faculty of Economics and Business in the University of
Murcia and in the Mathematics Department of  the University of Alicante.
 His research interests include information theory, coding theory and cryptography.
\end{IEEEbiographynophoto}

\begin{IEEEbiographynophoto}{J. J. Sim\'on}
(Juan Jacobo Sim\'on Pinero) received his degree in Mathematics in the UNAM of M\'exico in 1988 and its Ph.D. degree in Mathematics in the University of Murcia, Spain, in 1992.
Since 1995 he is Professor in the University of Murcia. Dr. Sim\'on has published papers in ring theory, group rings, partial actions of groups and coding theory.
\end{IEEEbiographynophoto}




\end{document}